\documentclass[fleqn,usenatbib]{mnras}

\usepackage{newtxtext,newtxmath}

\usepackage[T1]{fontenc}

\DeclareRobustCommand{\VAN}[3]{#2}
\let\VANthebibliography\thebibliography
\def\thebibliography{\DeclareRobustCommand{\VAN}[3]{##3}\VANthebibliography}

\usepackage{graphicx}	
\usepackage{amsmath}	
\usepackage[dvipsnames]{xcolor}
\usepackage{enumitem}

\usepackage{longtable}
\usepackage{ltablex}
\usepackage{caption, booktabs}
\usepackage{multicol}

\title[Disk chemistry with winds]{Impact of photoevaporative winds in chemical models of externally irradiated protoplanetary disks}

\author[L. Keyte et al.]{
Luke Keyte$^{1}$\thanks{E-mail: l.keyte@qmul.ac.uk} and 
Thomas J. Haworth,$^{1}$
\\
$^{1}$Astronomy Unit, School of Physics and Astronomy, Queen Mary University of London, London E1 4NS, UK\\
}

\date{Accepted 2025 January 7. Received 2025 January 3; in original form 2024 November 11.}

\pubyear{\the\year{}}

\begin{document}
\label{firstpage}
\pagerange{\pageref{firstpage}--\pageref{lastpage}}
\maketitle

\begin{abstract}
Most stars form in dense clusters within high-mass star-forming regions, where protoplanetary disks may be exposed to intense UV radiation from nearby massive stars. While previous studies have typically focused on isolated sources in low-mass regions, recent observational campaigns have started to probe the chemistry of irradiated disks in unprecedented detail. Interpreting this data requires complex chemical models, yet few studies have examined these disks' chemistry, and none have incorporated the photoevaporative wind launched by external UV fields into their physical structure. In this study, we post-process radiation hydrodynamics simulations of externally irradiated protoplanetary disks using the thermochemical code \textsc{dali}, comparing models with and without the wind to assess its impact on disk chemistry. Results show that UV radiation is rapidly attenuated by the disk in both cases. However, thermal re-emission from the wind at longer wavelengths enhances disk heating, increasing the gas-phase abundances of some key volatiles. Synthetic line fluxes vary by orders of magnitude between wind and windless models, primarily due to emission from the wind itself rather than abundance variations within the disk. Our findings demonstrate that the photoevaporative wind significantly influences the physical and chemical structure, and observational characteristics, of externally irradiated disks. We conclude that incorporating the wind into chemical models is essential for accurately predicting chemical abundances, interpreting observations, and ultimately understanding planet formation in these common yet complex environments.
\end{abstract}



\section{Introduction}

Planets form in protoplanetary disks surrounding young stars. Most disks are not isolated, but reside within stellar clusters in high-mass star-forming regions \citep[SFRs; e.g.][]{lada_lada_2003, kennicutt_evans_2012, krumholz_2019}, where they may be exposed to intense UV irradiation from nearby massive O- and B-type stars \citep[for a review see][]{winter_haworth_2022}. This external irradiation significantly influences disk evolution and, consequently, may influence the planet formation process \citep[e.g.][]{WinterPF, QiaoPF, HuangPF}. As the brightest and most easily resolved sources, most detailed studies of disk properties have focused on nearby, relatively isolated disks in low-mass SFRs \citep[e.g.][]{ansdell_2016, long_2018, DSHARPOverview, MAPSOverview}, or on those surrounding intermediate-mass Herbig Ae/Be stars \citep[e.g.][]{isella_2016, lazareff_2017, boehler_2018, keyte_2023, booth_2024b, booth_2024a}. While these disks offer valuable insights into planet formation processes, they may not represent the typical conditions under which planets form \citep[e.g.][]{FatuzzoAndAdams2008, WinterPrevalence, winter_haworth_2022}. Understanding how disks form and evolve in high-mass SFRs is crucial for building realistic planetary formation models that can accurately predict the formation pathways of planets in a variety of environments.

Existing studies of externally irradiated disks have primarily focused on quantifying fundamental properties such as mass, size, morphology, lifetime, and evolution, guided by both theoretical models and observational data \citep[e.g.][]{bally_1998b, adams_2004, vicente_alves_2005, facchini_2016, eisner_2006, van_terwisga_2019, boyden_eisner_2020, ColemanAndHaworth22, van_terwisga_2023, coleman_2024}. Observational evidence for externally irradiated disks has primarily come from spatially resolved Hubble Space Telescope observations of proplyds in the Orion Nebula Cluster \citep[ONC; e.g.][]{odell_wen_hu_1993, odell_wen_1994, mccaughrean_odell_1996, chen_young_1998, bally_1998a, bally_2000, Ricci_2008}. These observations demonstrate that external photoevaporation significantly reduces disk mass, size, and lifetime, with inferred mass-loss rates reaching up to $\sim 10^{-6} M_\odot$ yr$^{-1}$ in extreme cases \citep[e.g.][]{Henney_1999, facchini_2016, haworth_2018_fried, haworth_clarke_2019, 2021MNRAS.501.3502H, 2024A&A...687A..93A}. Complementary observations using the Atacama Large Millimeter/submillimeter Array (ALMA) at millimeter wavelengths have provided thermal dust continuum measurements, corroborating that externally irradiated disks are generally more compact and less massive than their isolated counterparts \citep[e.g.][]{mann_2014, eisner_2016, eisner_2018, ballering_2023, van_terwisga_2023}. While ALMA has also detected molecular line emission in a subset of these disks, their typical distance ($\sim 400$ pc) compared to well-studied isolated disks ($\lesssim 200$ pc) often poses challenges in spatially resolving the gas component \citep{boyden_eisner_2020, diaz_berrios_2024, goicoechea_2024}. Upcoming observational campaigns with ALMA and the James Webb Space Telescope (JWST) will provide the most detailed view of disk chemistry in irradiated environments to date, opening new avenues for understanding the impact of external photoevaporation on disk chemistry and, consequently, on planet formation in clustered environments.

To interpret these complex observational data, thermochemical models are essential. \citet{walsh_2013} conducted one of the first comprehensive analyses of disks subject to intense external irradiation, simulating a T-Tauri disk exposed to a $4\times 10^5$ G$_0$ UV field. Their models demonstrated significant effects of intense external irradiation on disk temperature structure and chemical composition, particularly pronounced in cold outer regions ($r\sim 100$ au). While these models predicted enhanced abundances of key volatiles such as CN, CS, and CO$_2$ - with disk-integrated fluxes typically increasing by factors of a few — the disk interior remained predominantly molecular due to efficient UV shielding, showing only minor variations within 10 au. However, recent observational studies have yielded conflicting results regarding these model predictions. ALMA observations of molecular lines in ONC disks 216-0939 and 253-1536A/B revealed chemical compositions remarkably similar to those of isolated disks in low-mass star-forming regions, based on both column density measurements and line flux ratios \citep{diaz_berrios_2024}. Similarly, JWST-MIRI observations of the irradiated disk XUE 1 indicate inner disk chemistry consistent with isolated systems \citep{ramirez_tannus_2023}. In contrast, other studies suggest potential variations in irradiated environments. For example, \citet{boyden_eisner_2023} found that irradiated disks in the ONC maintain gas-phase CO abundances close to ISM levels (CO/H $\sim 10^{-4}$), which they attribute to higher dust temperatures preventing CO freezeout. Similarly, an undepleted carbon reservoir has been inferred in the irradiated disk d203-506, where strong external UV radiation raises dust temperatures and enhances photodesorption rates \citep{goicoechea_2024}. These divergent findings highlight the complexity of planet-forming environments under external irradiation and underscore the need for refined thermochemical models that can reconcile these apparent contradictions.

To date, most chemical studies of irradiated disks have adopted a similar methodology, using 1D slab models \citep[e.g.][]{goicoechea_2024} or prescribing the disk density structure following the typical power-law surface density profile introduced by \citet{lyndenbell_pringle_1974}, and applying an isotropic background UV field. However, spatially resolved observations of irradiated sources reveal that the density structure becomes far more complex under these extreme conditions, with the intense UV field launching a photoevaporative wind from the disk surface. This is corroborated by theoretical and hydrodynamical simulations, which show that the wind can significantly alter the disk structure and mass distribution \citep[e.g.,][]{johnstone_1998, adams_2004, clarke_2007, haworth_2018_fried}. However, solving for the wind structure properly requires iteratively solving photodissociation region (PDR) chemistry and hydrodynamics. Multi-dimensional radiation hydrodynamic models are therefore computationally expensive. As such, combining hydrodynamics with complex disk chemistry is generally unfeasible. The impact of including the photoevaporative wind on the composition of the disk in chemical models of externally irradiated disks therefore remains unclear. Since the wind can present a significant column to the external radiation field, which makes the opacity structure of the source more complex and impacts the heating and cooling balance, we may reasonably expect that including the wind in chemical models is vitally important for accurately predicting disk composition and evolution.

This study asks the question: “How do the winds themselves influence externally irradiated disk chemistry?". We run a suite of chemical models for irradiated disks, comparing scenarios with and without the inclusion of the wind component. Our models span a range of UV field strengths to explore how the presence of a photoevaporative wind impacts disk chemistry across diverse environments. By contrasting the chemical compositions, molecular abundances, and emission line strengths predicted by wind and no-wind models, we aim to quantify the importance of including wind structures in chemical models of externally irradiated disks. Additionally, we discuss potential observational signatures and provide guidance for future observational campaigns with facilities such as ALMA.

This paper is organised as follows: our disk model is presented in Section 2, results are presented in Section 3, and their implications are discussed in Section 4. We present our conclusions and suggestions for future work in Section 5.

\section{Modelling}

\subsection{Overview}
At present there is no quick way to study the chemistry of externally irradiated disks that include a wind component. To make a first assessment of the impact of the wind, we utilise existing radiation hydrodynamic simulations from the literature to obtain detailed density structures of protoplanetary disks exposed to external UV fields ranging from 100 G$_0$ to 5000 G$_0$. The external UV field is defined as the integrated flux between 912-2400 \AA \; in Habing units (G$_0 \sim 1.6 \times 10^{-3}$ erg s$^{-1}$ cm$^{-2}$). These computationally intensive models provide density distributions which include both the disk and its associated photoevaporative wind. We then post-process these density structures using a state-of-the-art thermochemical code to determine the temperature structure, radiation field, and chemical abundances throughout the disk and wind regions.

To isolate the influence of the photoevaporative wind on disk chemistry, we conduct paired simulations for each scenario: one incorporating the full density structure from the hydrodynamical model, and another where the material in the wind region is artificially removed. This comparative approach allows us to quantify the significance of including the wind component in chemical models of externally irradiated protoplanetary disks, while maintaining computational feasibility. The following subsections provide detailed descriptions of the hydrodynamical models used as inputs (Section \ref{subsec:modelling_hydromodel}) and the thermochemical code used for the chemical analysis (Section \ref{subsec:modelling_chemicalmodel}).

\subsection{Hydrodynamic model}
\label{subsec:modelling_hydromodel}

We make use of existing hydrodynamical models first presented in \citet{ballabio_2023}. These models were generated using the photochemical hydrodynamics code \textsc{torus-3d-pdr} \citep{bisbas_2015}, and are similar to those presented in \citet{haworth_clarke_2019}. \textsc{torus-3dpdr} is a coupled code which combines the radiation hydrodynamics scheme from \textsc{torus} \citep{haworth_harries_2012} with the multi-dimensional radiative transfer, equilibrium chemistry and thermal balance calculations from the PDR code \textsc{3d-pdr} \citep{bisbas_2012}. For a detailed description of the modelling procedure, we direct readers to these studies. Here, we simply restate some of the key features.

\begin{figure*}
\centering
\includegraphics[clip=,width=1.0\linewidth]{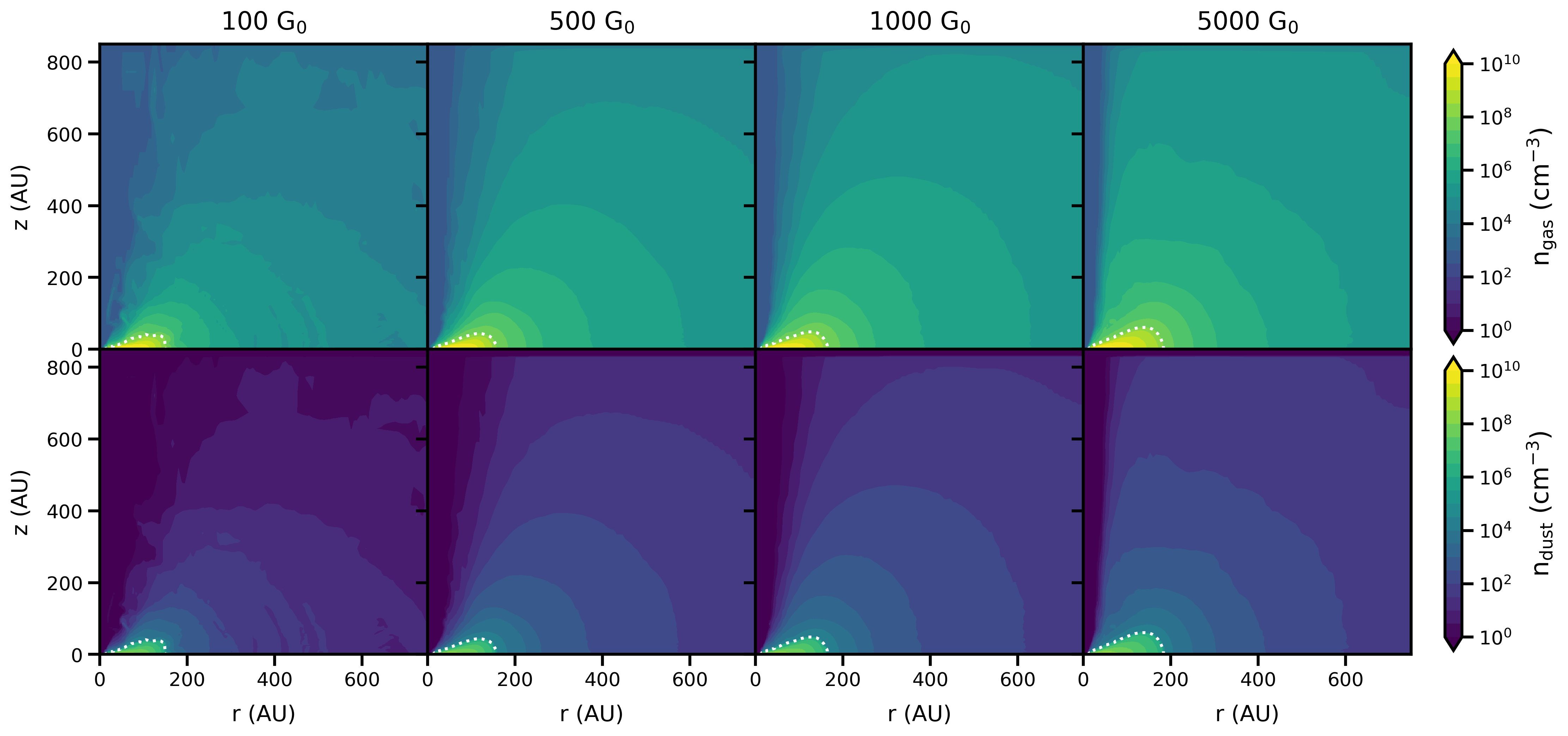}
\caption{Density structure for models including a photoevaporative wind. The dotted white line denotes the $n_\text{gas}=10^8$ contour, which we use as a cut-off point for our `no-wind' models, in which all gas and dust outside of this region is artificially removed. \emph{Top row: }Gas number densities for models with varying external UV fields (100 G$_0$, 500 G$_0$, 1000 G$_0$, 5000 G$_0$). \emph{Top row: }Corresponding dust number densities.}
\label{fig_density_wind}
\end{figure*}

The radiation hydrodynamic models use an imposed `disk' boundary condition, up to one scale height. This consists of a truncated power-law surface density profile
\begin{equation}
    \Sigma (r) = \Sigma_\text{1au} \; \bigg(\frac{r}{\text{au}}\bigg)^{-1}
\end{equation}
where $\Sigma_\text{1au}$ is the surface density at $r=1$ au, given by
\begin{equation}
     \Sigma_\text{1au} = \frac{M_\text{d}}{2 \pi r_\text{d} \text{au}}
\end{equation}
where the models we use all assume $M_\text{d} = 0.1 M_\odot$ for an $R_d=100$\,au disk outer boundary, i.e. $\Sigma_\text{1au} = 1407 $\,g\,cm$^{-2}$. The disk is imposed up to one scale height
\begin{equation}
     H = \frac{c_\text{s}}{\Omega}
\end{equation}
where $c_\text{s}$ is the sound speed and $\Omega$ the Keplerian angular velocity. The scale height determining the region up to which the disk is imposed as a boundary condition is computed assuming an isothermal disk temperature of 20 K \citep{haworth_clarke_2019}. 

The disk is irradiated passively by the central star and by an isotropic external UV field. The temperature due to stellar irradiation is assumed to follow the form
\begin{equation}
     T_* = 100 \; \bigg( \frac{r}{\text{au}} \bigg)^{-1/2}
\end{equation}
Four scenarios are considered, each using a different external UV field: 100 G$_0$, 500 G$_0$, 1000 G$_0$, and 5000 G$_0$. All other parameters remain constant across simulations. This range of UV field strengths represents low to intermediate levels of external irradiation of disks. For reference, these UV fields would correspond to distances ranging from around 5.5 to 0.75\,pc from an O star like $\theta^1$\,Ori\,C in the Orion Nebula Cluster and so are not representative of the strongly irradiated proplyds there \citep[though proplyds have been observed in NGC 1977 down to $\sim3000$\,G$_0$;][]{2016ApJ...826L..15K}. We note also that UV fields of order $10^3$ may represent the peak of the distribution of UV fields that disks are exposed to \citep{FatuzzoAndAdams2008, WinterPrevalence, winter_haworth_2022}. 

The disk beyond the boundary and in the wind is then allowed to evolve, with hydrodynamics and PDR calculations computed iteratively. For the radiation hydrodynamics a reduced UMIST 2012 \citep{mcelroy_2013_umist} chemical network (33 species, 330 reactions; \citealt{bisbas_2012}) is used, tailored to calculate accurate temperatures as efficiently as possible. It is important to note that this chemical network is used solely to accurately determine the physical structure; we perform more detailed modelling in a post-processing step to compute the temperature structure, radiation field, and chemical abundances (Section \ref{subsec:modelling_chemicalmodel}).

The gas and dust density structures derived from each of these models are presented in Figure \ref{fig_density_wind}. In each case, the model domain extends to $r=800$ au and $z=850$ au, and is dominated by the photoevaporative wind. These density structures serve as inputs to the chemical model, described in the next section.

\begin{figure*}
\centering
\includegraphics[clip=,width=1.0\linewidth]{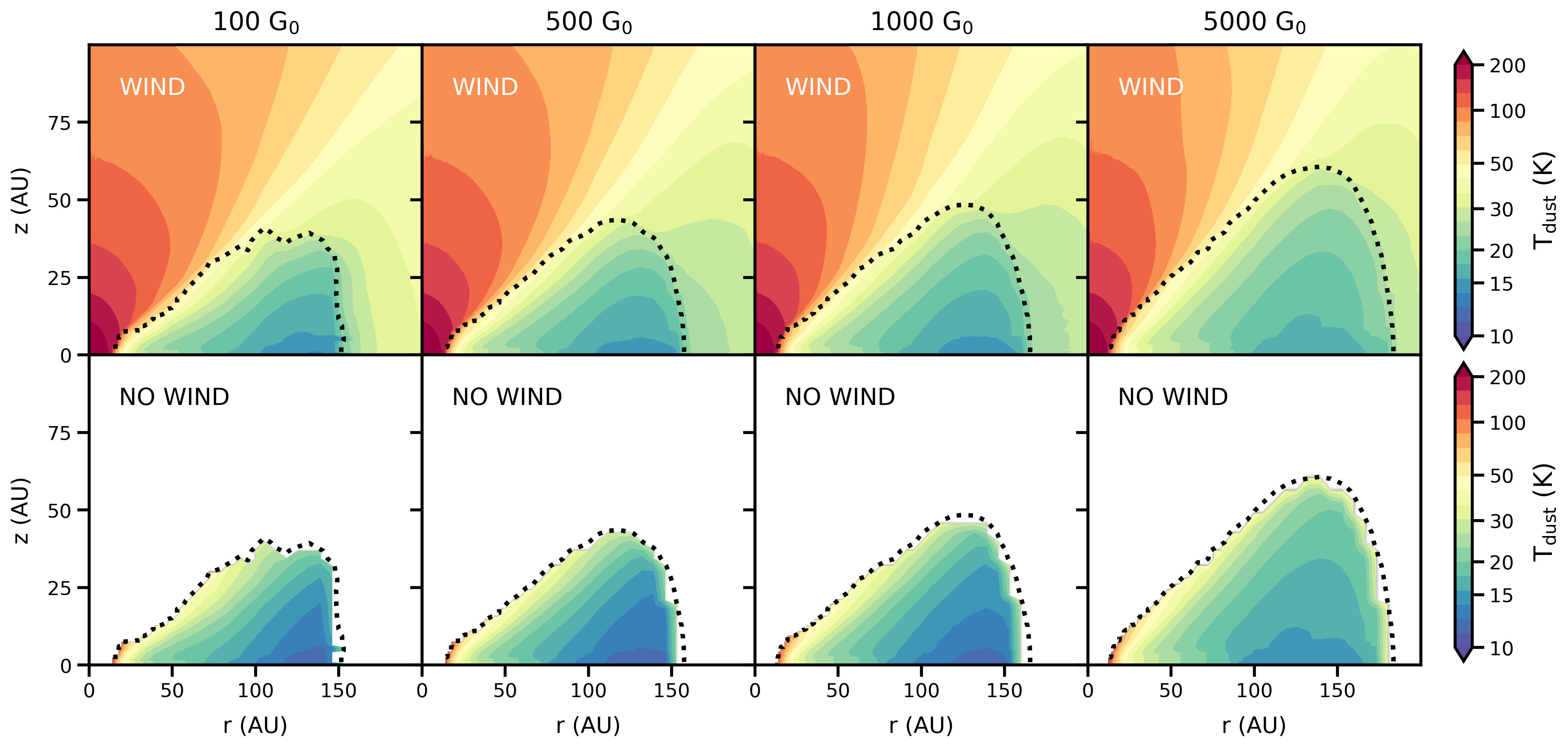}
\caption{Dust temperature structure for models including a photoevaporative wind (top row) and those without (bottom row). In each case, the dotted black line denotes the $n_\text{gas}=10^8\,$cm$^{-3}$ contour, which we use as a cut-off point for our `no wind' models, in which all gas and dust outside of this region is removed.}
\label{fig_tdust}
\end{figure*}

\subsection{Chemical model}
\label{subsec:modelling_chemicalmodel}
To determine the disk temperature structure, radiation field, and chemical abundances, we make use of the 2D thermochemical code \textsc{dali} \citep{Bruderer2012, Bruderer2013}. Each model run begins with a gas and dust density distribution structure from the hydrodynamical model outputs described previously. A Monte Carlo radiative transfer scheme is then used to determine the UV radiation field and dust temperature. The code includes two sources of UV photons; the host star, defined by an input stellar spectrum, and the background interstellar radiation field (ISRF), with photons propagated uniformly from a `virtual sphere' encompassing the model domain. The computed dust temperature serves as an initial estimate for the gas temperature, initiating an iterative process where the chemistry is solved time-dependently. Finally, a raytracing module is used to obtain disk-integrated line fluxes.

\subsubsection{Physical structure}
In order to make the chemical calculations computationally feasible, it is necessary to interpolate the gas density structure from the hydrodynamical models on to a coarser grid. Our new grid comprises 200 radial and 100 vertical cells, reducing the total cell count by a factor of $\sim 11$. Radially, we use 100 logarithmically-spaced cells out to 100 au, followed by 100 linearly-spaced cells extending to the model's outer boundary ($\sim$800 au). The vertical grid uses logarithmic spacing to provide better resolution near the midplane, which is necessary to account for settling of large dust grains.

We define an arbitrary density threshold to demarcate the disk from the wind. This enables us to compare scenarios with and with out the wind, and also defines where we transition to the wind having smaller maximum grain size and depleted dust-to-gas ratio  ($\Delta_\text{d/g}$), as expected theoretically \citep[][Paine et al. in prep]{facchini_2016, 2021MNRAS.508.2493O} and implemented in the underlying hydrodynamical model. We adopt a density threshold of $n_\text{gas}=10^8$\,cm$^{-3}$, setting $\Delta_\text{d/g}=0.01$ in the disk region and $\Delta_\text{d/g}=3\times 10^{-4}$ in the wind region. Wind dust consists of a single small grain population, with sizes $0.005\mu$m-$1\mu$m that follow the standard ISM grain size distribution \citep{mathis_1977}. The disk region incorporates two grain populations; small grains identical to those in the wind, and large grains ($0.005\mu$m-$1$mm) following the same size distribution. 

Dust settling is parametrised using $f_0$ and $\chi$, representing the fraction of total dust mass in large grains at the midplane and the fraction of scale height to which large grains settle, respectively. The large-to-small grain ratio decreases with height according to:
\begin{equation}
    f = f_0 \exp \bigg[ -\frac{1}{2} \bigg( \frac{\pi/2 - \theta}{\chi h} \bigg) ^{2} \bigg]
\end{equation}
where $h$ is the scale height angle and $\pi/2-\theta \sim z/r$. We adopt typical values of $f_0=0.9$ and $\chi=0.2$ for all models \citep[e.g.][]{kama_2016a, fedele_2017, trapman_2020b, leemker_2022, keyte_2024b, keyte_2024a, stapper_2024}.

\subsubsection{Stellar parameters}
The stellar spectrum is modelled as a blackbody with temperature $T = 5500$ K  and luminosity $L_* = 1 \; L_\odot$. The X-ray spectrum was characterised as a thermal spectrum with a temperature of $7 \times 10^7$ K over the 1 to 100 keV energy range, with an X-ray luminosity of $L_X = 7.94 \times 10^{28}$ erg s$^{-1}$.

\begin{figure*}
\centering
\includegraphics[clip=,width=1.0\linewidth]{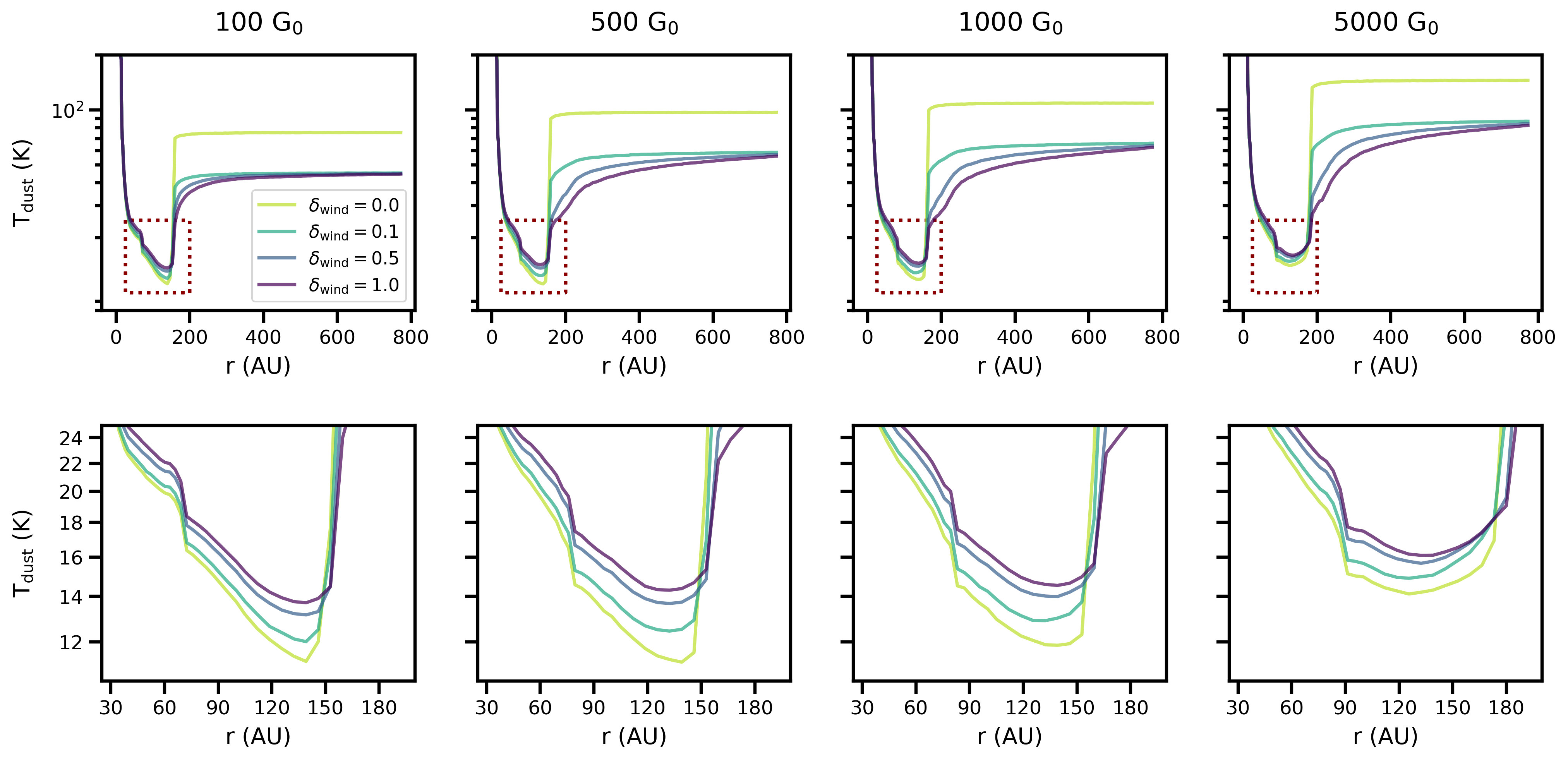}
\caption{\emph{Top row: }Midplane dust temperatures for models with an external UV field between 100 to 5000 G$_0$. Coloured lines denote different wind depletion factors (where$\delta_\text{wind} = 0$ corresponds to the wind being entirely removed and $\delta_\text{wind}=1$ corresponds to no gas or dust removed). \emph{Bottom row: }Zoom in of the region bounded by the red box in top panel, highlighting the increased dust temperatures at the midplane when including a photoevaporative wind.}
\label{fig_temp_midplane}
\end{figure*}

\begin{figure*}
\centering
\includegraphics[clip=,width=1.0\linewidth]{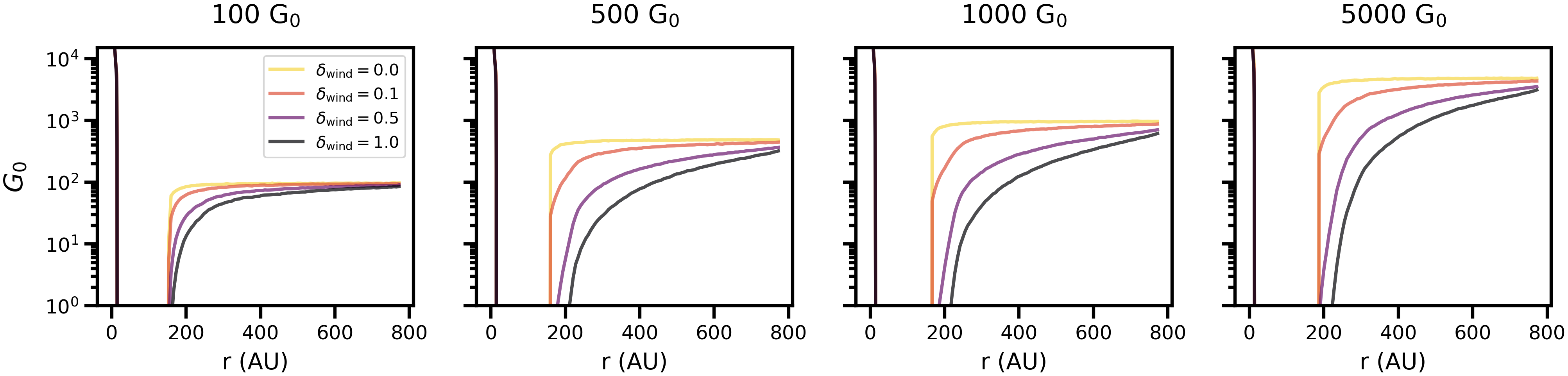}
\caption{Midplane integrated FUV flux (in units of the Draine field, $G_0$) for models with an external UV field between 100 to 5000 G$_0$. Coloured lines denote different wind depletion factors (where $\delta_\text{wind} = 0$ corresponds to the wind being entirely removed and $\delta_\text{wind}=1$ corresponds to no gas or dust removed).}
\label{fig_g0_midplane}
\end{figure*}

\subsubsection{Chemical network}
We model the chemistry using a network based on a subset of the UMIST 06 network \citep{woodall2007}. This network was first presented in \citet{visser_2018} and includes 134 species and 1855 reactions. The code includes H$_2$ formation on dust, freeze-out, thermal desorption, hydrogenation, gas-phase reactions, photodissociation and photoionisation, X-ray induced processes, cosmic-ray induced reactions, PAH charge exchange/hydrogenation, and reactions with vibrationally excited H$_2$. Non-thermal desorption is only included for a small number of species (CO, CO$_2$, H$_2$O, CH$_4$, NH$_3$, N$_2$). For grain surface chemistry, only hydrogenation of simple species is considered (C, CH, CH$_2$, CH$_3$, N, NH, NH$_2$, O, OH, CN). The details of these processes are described more fully in \citet{Bruderer2012}.

As inputs to the chemical network, we adopt initial molecular abundances based on \citet{ballering_2021} (see Table \ref{table:modelparameters}). The chemistry is evolved for 1 Myr, typical of the age of the types of system investigated in this work.

\section{Results}

\subsection{Temperature structure and radiation field}
\label{subsec:results_temperature}

Figure \ref{fig_tdust} shows the temperature profiles for each of our models, with interstellar radiation field intensities ranging from 100 to 5000 G$_0$. For each radiation field strength, we compare two cases; models that include a photoevaporative wind (top panels) and those without (bottom panels). Our analysis reveals that within the disk itself — excluding the wind region — temperatures are consistently several Kelvin higher in models with photoevaporative winds compared to those without, regardless of the external radiation field strength.

To further investigate this relationship, we conducted additional simulations wherein the wind density was systematically reduced by a factor $\delta_\text{wind}$. In this context, $\delta_\text{wind}=0$ represents complete wind removal, while $\delta_\text{wind}=1$ represents unaltered wind density. We computed dust temperatures for $\delta_\text{wind}$ values of 0, 0.1, 0.5, and 1, with results presented in Figure \ref{fig_temp_midplane}. The upper row displays the midplane dust temperature as a function of $\delta_\text{wind}$ for each model, while the lower row focuses on the $r = 30-200$ au region, where temperature variations are most pronounced.

Across all models, a progressive reduction in wind density corresponds to a gradual decrease in midplane dust temperature. While absolute temperature values vary with background radiation field intensity — higher UV intensities generally yielding warmer conditions — the relative temperature changes due to wind depletion remain broadly consistent across models. The maximum temperature differential between full-wind and no-wind scenarios is approximately $\Delta T \approx 3$ K. 

The midplane FUV flux for each of these models is illustrated in Figure \ref{fig_g0_midplane}. Within the wind itself, the FUV flux is strongly dependent on the wind depletion factor $\delta_\text{wind}$, with the flux becoming attenuated by up to $\sim 2$ orders of magnitude comparing the full-wind and no-wind models. However, regardless of the background UV field intensity or the wind depletion factor, the UV flux undergoes rapid attenuation upon reach the disk/wind boundary. Consequently, all models have midplane FUV flux values within the disk itself that are significantly below 1 G$_0$. This consistent attenuation occurs irrespective of variations in the external radiation environment or wind properties. The fact that the midplane FUV field strength within the disk is negligible for all values of $\delta_\text{wind}$ suggests that the additional heating observed in models including the wind is not driven directly by FUV radiative heating. We explore this phenomenon in greater detail in Section \ref{sec:discussion}.

\subsection{Chemical composition}
\label{subsec:results_composition}

\begin{figure*}
\centering
\includegraphics[clip=,width=1.0\linewidth]{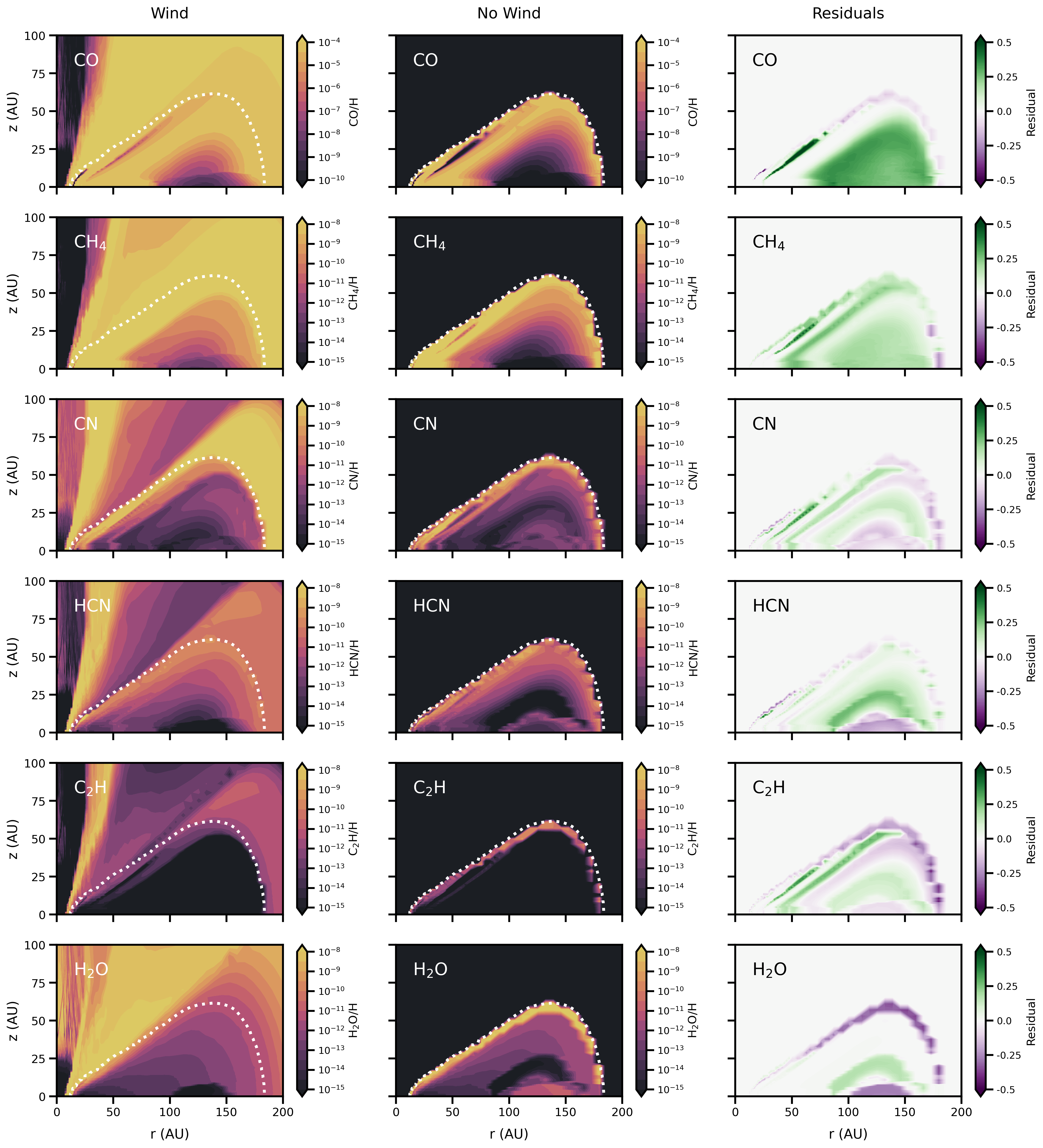}
\caption{Abundance maps for key volatile species (CO, CH$_4$, CN, HCN, C$_2$H, H$_2$O) extracted from the model with a background UV field of 5000 G$_0$. The left column shows the full-wind scenario, the middle column shows the no-wind scenario, and the right column depicts the residual differences between the two cases. The residuals are calculated as the normalised logarithmic difference in abundance following Equation \ref{eq:residuals}. Green represents an enhancement in the wind model compared to the no-wind model, while purple represents depletion. Note that the CO abundance map uses a different color scaling to the other molecules in order to best showcase the dynamic range.}
\label{fig_abu_maps}
\end{figure*}

\begin{figure*}
\centering
\includegraphics[clip=,width=1.0\linewidth]{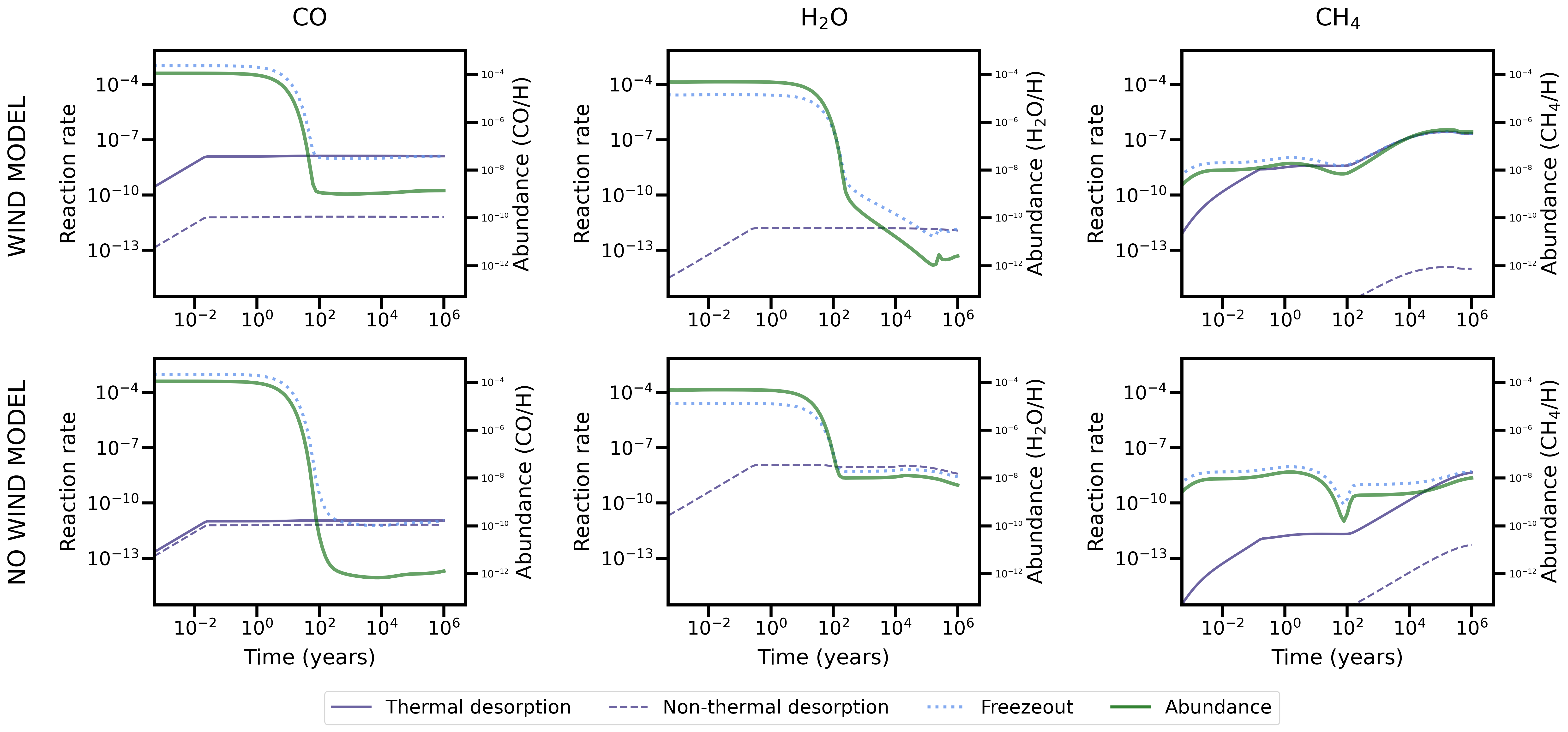}
\caption{CO, H$_2$O, and CH$_4$ freezeout/desorption rates (primary y-axis) and abundance variations (secondary y-axis) as a function of time for models with an external UV field of 5000 G$_0$. For each species, values were extracted from a characteristic grid cell where abundance variations are most pronounced (see main text). \emph{Top row: }Models including the photoevaporative wind. \emph{Bottom row: }Models where the photoevaporative wind has been removed.}
\label{fig_freezout_desorption}
\end{figure*}

This section investigates the impact of the wind on the chemical composition of the disk. To begin, we consider only the chemistry within the disk (defined as the region with densities $n_\text{gas} > 10^8$ cm$^{-3}$; Section \ref{subsec:modelling_chemicalmodel}), and ignore the composition of the wind itself. For simplicity, we focus on a select group of key volatile species: CO, CH$_4$, CN, HCN, C$_2$H, and H$_2$O. Figure \ref{fig_abu_maps} shows abundance maps for these species, extracted from our model with a background UV field of 5000 G$_0$, comparing full-wind and no-wind scenarios. To highlight the differences between these two cases, we also present residual maps calculated using:
\begin{equation}
    \frac{\log_{10}{(\text{[X/H]}_\text{wind})} - \log_{10}{(\text{[X/H]}_\text{no-wind})}}{|\max(\log_{10}(\text{[X/H]}_\text{wind}), \log_{10}(\text{[X/H]}_\text{no-wind}))|}
    \label{eq:residuals}
\end{equation}
where [X/H] represents the abundance of species X relative to hydrogen. The denominator term serves to normalise the residuals, ensuring they are proportional to the highest abundance value observed in either the full-wind or no-wind model. This approach allows for a more meaningful comparison of abundance variations across different species and regions of the disk.

Our results reveal that the photoevaporative wind's impact on chemical abundance varies significantly among species:

\begin{itemize}[leftmargin=*, align=left]
    \item CO exhibits substantial enhancement, from CO/H $\sim 10^{-12}$ to $10^{-9}$)
    towards the midplane between $r \sim 100$ to $150$ au in the full-wind model. However, this enhanced abundance remains significantly lower than in the inner disk and warm molecular layer (CO/H $\sim 10^{-4}$).
    \item CH$_4$ shows notable enhancement in the wind model, particularly in the upper disk regions close to the disk/wind boundary.
    \item CN, HCN, and C$_2$H show complex patterns of enhancement and depletion across well defined layers, but the absolute abundances remain relatively low throughout most of the disk.
    \item H$_2$O displays significant depletion in the upper disk atmosphere when the photoevaporative wind is included.
\end{itemize}

To understand the mechanisms driving these variations, we examined the dominant formation and destruction pathways for each species over time. Our analysis reveals that the observed differences are primarily driven by the balance between thermal desorption, non-thermal desorption, and freeze-out, which are influenced by the differing gas/dust temperatures and UV fields between the two models. Differences between other gas-phase formation and destruction routes for each species are comparatively minor (see Appendix \ref{apx:reaction_rates_analysis}).

Figure \ref{fig_freezout_desorption} (left column) illustrates this phenomenon for CO at $r = 125$ au, where the abundance disparity between the two models is most pronounced. In the full-wind model (upper panel), thermal desorption is the primary CO formation route. The freeze-out rate decreases with time until it equilibrates with thermal desorption at $\sim 10^{-8}$ s$^{-1}$, resulting in a steady-state abundance of CO/H $\sim 10^{-9}$. In contrast, the no-wind model experiences more extensive CO depletion due to lower dust temperatures. Here, the thermal desorption rate is lower ($\sim 10^{-10}$ s$^{-1}$), and the freeze-out rate eventually decreases to this level, leading to a significantly lower equilibrium abundance (CO/H $\sim 10^{-12}$).

This behaviour is not mirrored in the midplane abundances of other species due to their higher binding energies compared to CO. Although the temperature difference between the two models is only $\sim 3$ K, it is sufficient to dramatically shift the balance between thermal desorption and freeze-out for CO while leaving other species largely unaffected. The higher binding energies of other molecules mean they remain predominantly frozen onto grain surfaces under both wind and no-wind conditions in the outer disk, requiring higher temperatures to exhibit similar gas-phase abundance enhancements.

In the disk atmosphere, both models show rapid attenuation of the background UV field. However, the no-wind model maintains a significantly higher UV field in a radially extended, vertically narrow region compared to the full-wind model, which plays a crucial role in setting the gas-phase H$_2$O abundance in this region. The higher UV field in the no-wind model drives increased non-thermal desorption rates for H$_2$O (Figure \ref{fig_freezout_desorption}, middle column), while thermal desorption remains negligible. Conversely, for CH$_4$, which has a relatively low binding energy, the increased thermal desorption in the full-wind model dominates over the enhanced non-thermal desorption in the no-wind model, resulting in higher CH$_4$ abundances in the full-wind scenario (Figure \ref{fig_freezout_desorption}, right column). This chemistry is analysed in more detail in Appendix \ref{apx:reaction_rates_analysis}.

While the differences between the wind and no-wind models drive abundance variations spanning up to three orders of magnitude for some species, the absolute abundances resulting from these enhancements or depletions remain relatively low compared to other regions of the disk. Consequently, these localised variations likely do not significantly impact the total abundance of each species integrated over the entire disk. This suggests that while the presence of a photoevaporative wind in the model can induce notable chemical changes in specific disk regions, its overall effect on the disk-averaged molecular abundances may be limited for disks exposed to UV fields below 5000 G$_0$, as modelled here. We caution that these conclusions may not extend to more highly irradiated disks, where stronger UV fields could potentially drive more substantial chemical changes. We discuss this in more detail in Section \ref{subsec:impact_of_params}.

\begin{figure*}
\centering
\includegraphics[clip=,width=0.95\linewidth]{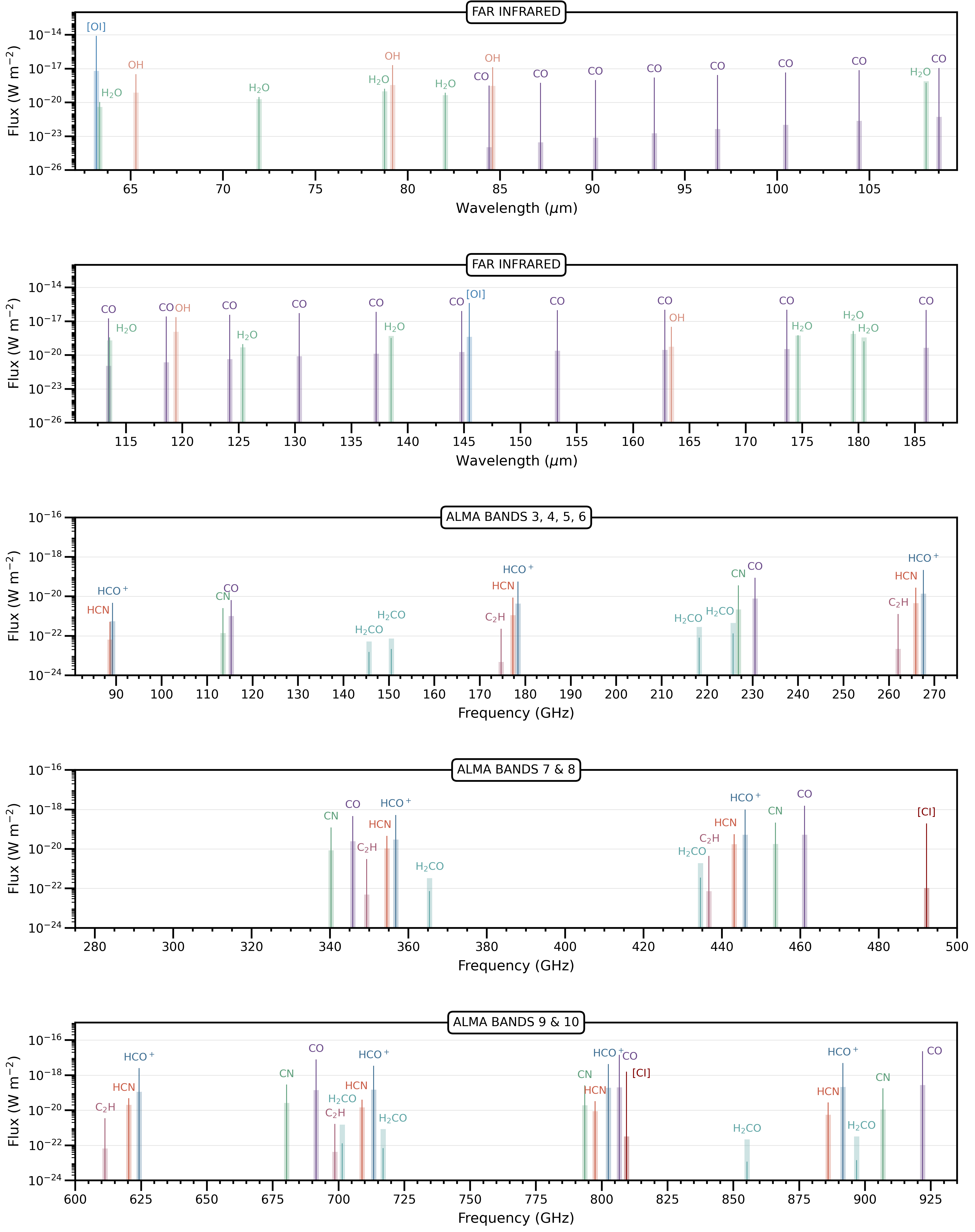}
\caption{Disk-integrated line fluxes from the full-wind model (thin lines) and no-wind model (thick lines), for the case where the background UV intensity is 5000 G$_0$. The far infrared fluxes (upper panels) cover a wavelength range similar to Herschel/PACS, while the remaining panels cover frequency ranges across ALMA Bands 3 to 10. For clarity, not all lines in Table \ref{table:line_fluxes} are shown.}
\label{fig_fluxes}
\end{figure*}

\subsection{Synthetic observations}
\label{subsec:synthetic_obs}

To evaluate the observational implications of incorporating a photoevaporative wind in our models, we generated synthetic disk-integrated line fluxes for a hypothetical face-on disk located at a distance of 400 pc, which is comparable to the distance to well-studied systems in the Orion Nebula Cluster \citep[e.g.][]{hillenbrand_1998}. Our analysis focused on atomic and molecular transitions observable at (sub-)millimeter wavelengths using ALMA, including CO, CN, HCN, [CI], HCO$^+$, C$_2$H, and H$_2$CO. Additionally, we considered a range of CO, H$_2$O, OH, and [OI] transitions in the far-infrared, which were previously accessible with the Herschel Space Observatory \citep{pilbratt_2010_herschel} and will be observable with the potential future NASA mission PRIMA (PRobe for-Infrared Mission for Astrophysics; \citealt{moullet_2023_prima}). We focus on models with an external UV field of 5000 $G_0$, comparing two scenarios: a full-wind model and one in which the wind has been artificially removed.
 
Figure \ref{fig_fluxes} presents the synthetic line fluxes for each atomic/molecular transition across both scenarios. Our results reveal substantial variations in line fluxes between the full-wind an no-wind models. All species, except for H$_2$CO, exhibit higher fluxes when the wind is included. The magnitude of these variations generally increases with the rotational quantum number $J$, indicating a more pronounced effect on higher frequency transitions that trace warmer gas. The flux increase when including the wind ranges from a factor of a few to greater than $10^5$ in the most extreme cases. The full list of synthetic line fluxes ($F_\text{wind}$ and $F_\text{no-wind}$) and corresponding line flux ratios $F_\text{wind}$/$F_\text{no-wind}$ are presented in Table \ref{table:line_fluxes}.

To understand the origin of these variations we analysed the line contribution function for each transition, which show the region of the disk from which a given fraction of the emission originates. Figure \ref{fig_cbfs} presents a representative example transition for each species in our analysis. Each panel displays the species abundance, overlaid with contours denoting the regions where 20\%, 40\%, and 60\% of the emission originates, comparing the line contribution functions for both full-wind and no-wind models. These examples demonstrate that the enhanced line fluxes in the full-wind model primarily result from significant contributions from the wind, rather than increased emission due to abundance enhancements within the disk itself, as described in the previous subsection. The wind region traced by the emission varies considerably depending on both the species and transition. 

These findings are consistent with the results presented in Section \ref{subsec:results_composition}, where we demonstrated that, although the inclusion of a wind in the chemical model can enhance or deplete the abundance of various molecular species within specific regions of the disk, the absolute abundances in those regions remain comparatively low. Consequently, the enhanced line fluxes are not directly tied to an increase in emission from the disk itself. The observational implications of these results are discussed in the next section.

\section{Discussion}
\label{sec:discussion}

\subsection{Significance of the photoevaporative wind}
\label{subsec:discussion_significance_of_wind}

We have demonstrated that the inclusion of the photoevaporative wind into the density structure of chemical models for externally irradiated protoplanetary disks leads to notable variations in the temperature profile, molecular abundances, and predicted line fluxes. These findings both complement and challenge our existing understanding of chemistry in externally irradiated environments.

A key result from our work is that the presence of a photoevaporative wind can elevate disk dust temperatures by several Kelvin, even in regions well-shielded from direct UV radiation. This temperature enhancement, while modest, influences the gas-phase abundances of volatile species, particularly those with low binding energies such as CO. This mechanism represents an important consideration absent from previous studies of externally irradiated disks that neglect the wind's contribution to density structure. While earlier research by \citet{walsh_2013} established that external irradiation can enhance disk temperatures, particularly in cold outer regions, our analysis identifies an additional heating pathway: wind-driven re-emission can warm the disk even when direct UV heating is negligible. This mechanism facilitates heating deeper in to the disk compared to direct UV photon heating, resulting in temperature enhancements extending to the midplane and inward to approximately 30 au in our models.

The impact of external irradiation on molecular abundances is more nuanced. \citet{walsh_2013} predicted that exposing a disk to an external UV field leads to significant abundance variations, with the column densities of some species becoming enhanced (CN, CS, CO$_2$) and others becoming depleted (N$_2$H$^+$, H$_2$O, C$_2$H$_2$). Our windless models with varying UV field strengths broadly confirm these predictions, showing similar magnitude variations in column densities (Fig \ref{fig_column_densities}). However, the inclusion of the photoevaporative wind introduces additional complexity. While the wind-induced temperature increase is modest ($\sim3$ K), it sufficiently alters the desorption-freezeout balance, enhancing the gas-phase abundance of several key volatiles by a few orders of magnitude within the disk. However, these enhancements predominantly occur in cold outer regions where baseline gas-phase abundances are minimal. Consequently, absolute abundances of wind-enhanced gas-phase species remain substantially below those in warmer inner regions where they naturally predominate.

\begin{figure*}
\centering
\includegraphics[clip=,width=1.0\linewidth]{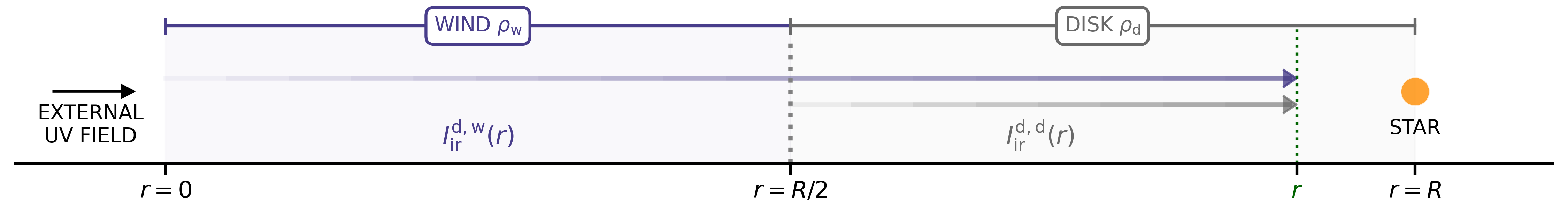}
\caption{Schematic illustrating the geometry of our 1D radiative transfer model. The model comprises two equally sized regions, `disk' and `wind', each characterised by a uniform density ($\rho_\text{d}$ and $\rho_\text{w}$). An external UV field impinges on the outer edge of the wind ($r=0$) and propagates inwards towards the star. The UV radiation is absorbed by the both the wind and disk, then remitted at IR wavelengths. The intensity of the IR radiation at a given point $r$ within the disk is determined by integrating the IR emission contributions from the wind and disk from $r=0$ to $r=r$.}
\label{fig_rt_schematic}
\end{figure*}
More dramatically, our models predict that including the wind substantially impacts observable line fluxes, with enhancements ranging from factors of a few to $\sim10^5$. These effects strengthen at higher frequencies, consistent with emission primarily originating from higher-temperature gas in the wind. Notably, flux variations between wind-inclusive and wind-free models frequently exceed variations between models differing only in UV field strength. The most significant enhancements appear in higher-$J$ CO rotational transitions (with successive transitions showing progressively larger enhancements) and atomic species (O and C showing $\gtrsim 10^3$ flux increases). However, it is important to note that these predictions represent an idealized scenario: a face-on disk with an isotropic UV field and axisymmetric wind. Real systems, with varying inclinations and localized winds, likely exhibit more modest flux enhancements.

Recent observational studies provide important context for these theoretical predictions. \citet{boyden_eisner_2023} conducted a comprehensive ALMA study of 20 irradiated disks in the ONC, revealing distinctive characteristics including high disk masses ($\gtrsim 10^{-3} - 10^{-2} M_\odot$), elevated gas-to-dust ratios ($\sim 100 - 1000$), and ISM-like CO abundances, contrasting with the carbon depletion often observed in isolated systems \citep[e.g.][]{kama_2016b, bosman_2021_maps}. Their analysis was based on thermochemical modelling CO $J=3-2$ and HCO$^+$ $J=4-3$ emission from hydrostatic windless disks. In our 5000\,G$_0$ model these lines both show an equal enhancement of a factor $\sim18$ in the wind case, which could conceivably contribute to the conclusion of the disks being gas rich. This comes with the caveat that our models do not extend to the same strength of FUV radiation field as the proplyds studied by \citet{boyden_eisner_2023}. The enhancement in the fluxes in the wind case summarised in Table \ref{table:line_fluxes} could provide useful diagnostic power. For example, low-lying CO rotational transitions, which show minimal wind enhancement could be combined with millimeter transitions of C$_2$H (typically enhanced by a factor $\sim 50$ when accounting for the wind) to identify potential wind signatures. High-$J$ CO transitions (J$\gtrsim 6$) similarly show $\gtrsim 50$ factor enhancements in wind-inclusive models.

However, some recent observational studies suggest that irradiated disk chemistry may not differ significantly from isolated disks. \citet{diaz_berrios_2024} detected various molecular species (CO, HCN, H$_2$CO, C$_2$H) in two ONC disks but found flux ratios comparable to isolated disks, contradicting previous model predictions of enhanced molecular emission. Similarly, \citet{ramirez_tannus_2023} demonstrated that the chemistry of the irradiated disk XUE 1 exhibited characteristics similar to isolated systems. They proposed that external irradiation truncates the disk's outer regions, which reduces the line-emitting region and results in molecular fluxes comparable to those observed in isolated disks.

Collectively, these findings emphasize that accurately modelling and interpreting observations of externally irradiated disks requires careful consideration of the wind's influence on both disk structure and observational diagnostics. The wind's impact manifests through multiple channels: direct modification of density structure, subtle but significant temperature changes affecting chemical abundances, and substantial contributions to molecular line emission. The challenge of discriminating between disk and wind emission is exacerbated by the typically compact nature and greater distances of irradiated disks compared to well-studied isolated systems. In the absence of spatially resolved observations delineating disk and wind components, future observational studies would benefit from multi-transition molecular line observations capable of differentiating these components, particularly through analysis of transitions with varying excitation temperatures that preferentially trace different regions of the disk-wind system.

\subsection{Origin of the wind-induced temperature enhancement}
\label{subsec:discussion_origin_of_wind}
In Section \ref{subsec:results_temperature}, we demonstrated that incorporating the photoevaporative wind into the chemical model's density structure results in enhanced heating of the disk. Although this enhancement is typically small ($\sim 3$ K), it can impact the gas-phase abundances of several key volatiles.

The mechanism behind this temperature enhancement is not immediately apparent. Our models reveal that UV radiation is rapidly attenuated by the disk, resulting in nearly identical UV flux at the midplane and within the warm molecular layer in both wind and windless cases. Thus, direct heating of dust grains by incident UV photons does not drive this enhancement. In fact, we might reasonably expect the photoevaporative wind to contribute to UV attenuation, potentially making the wind model colder than the windless model. However, this is not the case. Instead, we find that thermal re-emission from the wind drives the temperature enhancement. The wind absorbs UV photons and re-radiates predominantly in the infrared. Dust opacities are significantly lower at these longer wavelengths, allowing the re-emitted radiation to penetrate deeper into the disk and contribute to dust heating.

To better understand this phenomenon, we developed a simple 1D analytical radiative transfer model for a protoplanetary disk that includes a photoevaporative wind. This simplified frame work enables systematic examination of UV field reprocessing by the wind and wavelength-dependent photon absorption within the disk structure.

\subsubsection{Model setup}

Our  model geometry extends from  $r = 0$ to $r = R$, with two distinct regions: a wind region extending from $r = 0$ to $r = R/2$ characterized by uniform dust density $\rho_w$, and a disk region from $r = R/2$ to $r = R$ characterized by dust density $\rho_d$ (see Figure \ref{fig_rt_schematic}). This `inside-out' configuration sets the outer edge of the wind as the starting point, allowing us to consider how external radiation propagates inwards towards the star. The radiation field consists of an external UV field, $I_\text{uv}(0)$, irradiating the system from $r = 0$. The host star radiation field is not included. We make the assumption that, as UV radiation propagates inward, it is absorbed by dust grains and re-emitted entirely infrared (IR) radiation, maintaining energy conservation such that $E_\text{in} = E_\text{out}$. While this is a somewhat unrealistic over-simplification, it allows us the qualitatively investigate the manner is which the externally UV is reprocessed within the disk. Further, our model neglects scattering and considers only absorption processes. The absorption coefficients in the UV and IR are denoted as $\kappa_\text{uv}$ and $\kappa_\text{ir}$, respectively.

\subsubsection{UV absorption and IR re-emission}

In the wind region, the UV intensity at a distance $r$ is given by
\begin{equation}
    I_\text{uv}^w (r) = I_\text{uv}(0) \, e^{-\kappa_\text{uv} \rho_w r}
\end{equation}

\noindent In the disk region, the UV intensity at $r$ is
\begin{align}
    I^d_\text{uv} (r) &= I_\text{uv} (R/2) e^{-\kappa_\text{uv} \rho_d (r-R/2)} \\
                      &= I_\text{uv} (0) e^{-\kappa_\text{uv} \rho_w R/2} e^{-\kappa_\text{uv} \rho_d (r-R/2)}
\end{align}
Assuming thermal equilibrium, the dust in each region re-emits the absorbed UV energy as IR radiation. The emission coefficients in the wind and disk regions are
\begin{align}
    j_\text{ir}^w (r) &= \kappa_\text{uv} \rho_w  I_\text{uv}^w(r) \\
                      &= \kappa_\text{uv} \rho_w  I_\text{uv}(0) e^{-\kappa_\text{uv} \rho_w r}
\end{align}
\begin{align}
    j_\text{ir}^d (r) &= \kappa_\text{uv} \rho_d  I_\text{uv}^d(r) \\
                      &= \kappa_\text{uv} \rho_d  I_\text{uv}(0) e^{-\kappa_\text{uv} \rho_w R/2} e^{-\kappa_\text{uv} \rho_d (r - R/2)}
\end{align}

\noindent The IR intensity at a point $r$ in the disk region has two contributions: one from IR emission emanating from the wind region and another from the disk region itself. The contribution from the wind region is
\begin{align}
    I_\text{ir}^{d,w} (r) &= \kappa_\text{uv} \rho_w I_\text{uv}(0) \, e^{-\kappa_\text{ir} \rho_d (r - R/2)} \notag \\
                          & \times \int_0^{R/2} e^{-\kappa_\text{uv}\rho_w r'} e^{-\kappa_\text{ir} \rho_w (R/2 - r')} \, dr'
\end{align}

\noindent{while the contribution from the disk region is}
\begin{align}
    I_\text{ir}^{d,d} (r) &= \kappa_\text{uv} \rho_d I_\text{uv}(0) \, e^{-\kappa_\text{uv}\rho_w (R/2)} \notag \\
                          & \times \int_{R/2}^r e^{-\kappa_\text{uv}\rho_d (r' - R/2)} e^{-\kappa_\text{ir} \rho_d (r - r')} \, dr'
\end{align}

\noindent The total IR intensity at a point $r$ in the disk region is the sum of these contributions:
\begin{equation}
    I_\text{ir}^d (r) = I_\text{ir}^{d,w} (r) + I_\text{ir}^{d,d} (r)
\end{equation}

\noindent Thus
\begin{align}
    I_\text{ir}^{d} (r) &= \bigg[\kappa_\text{uv} \rho_w I_\text{uv}(0) \, e^{-\kappa_\text{ir} \rho_d (r - R/2)} \notag \\
                          & \times \int_0^{R/2} e^{-\kappa_\text{uv}\rho_w r'} e^{-\kappa_\text{ir} \rho_w (R/2 - r')} \, dr' \bigg] \notag\\
                          &+ \bigg[\kappa_\text{uv} \rho_d I_\text{uv}(0) \, e^{-\kappa_\text{uv}\rho_w (R/2)} \notag \\
                          & \times \int_{R/2}^r e^{-\kappa_\text{uv}\rho_d (r' - R/2)} e^{-\kappa_\text{ir} \rho_d (r - r')} \, dr' \bigg] \label{eq:ir_disk}
\end{align}
\noindent Using this expression, we investigate how variations in the wind and disk dust densities affect the propagation of IR radiation from the wind into the deeper regions of the disk.

\subsubsection{Application and parameter study}
\begin{figure}
\centering
\includegraphics[clip=,width=1.0\linewidth]{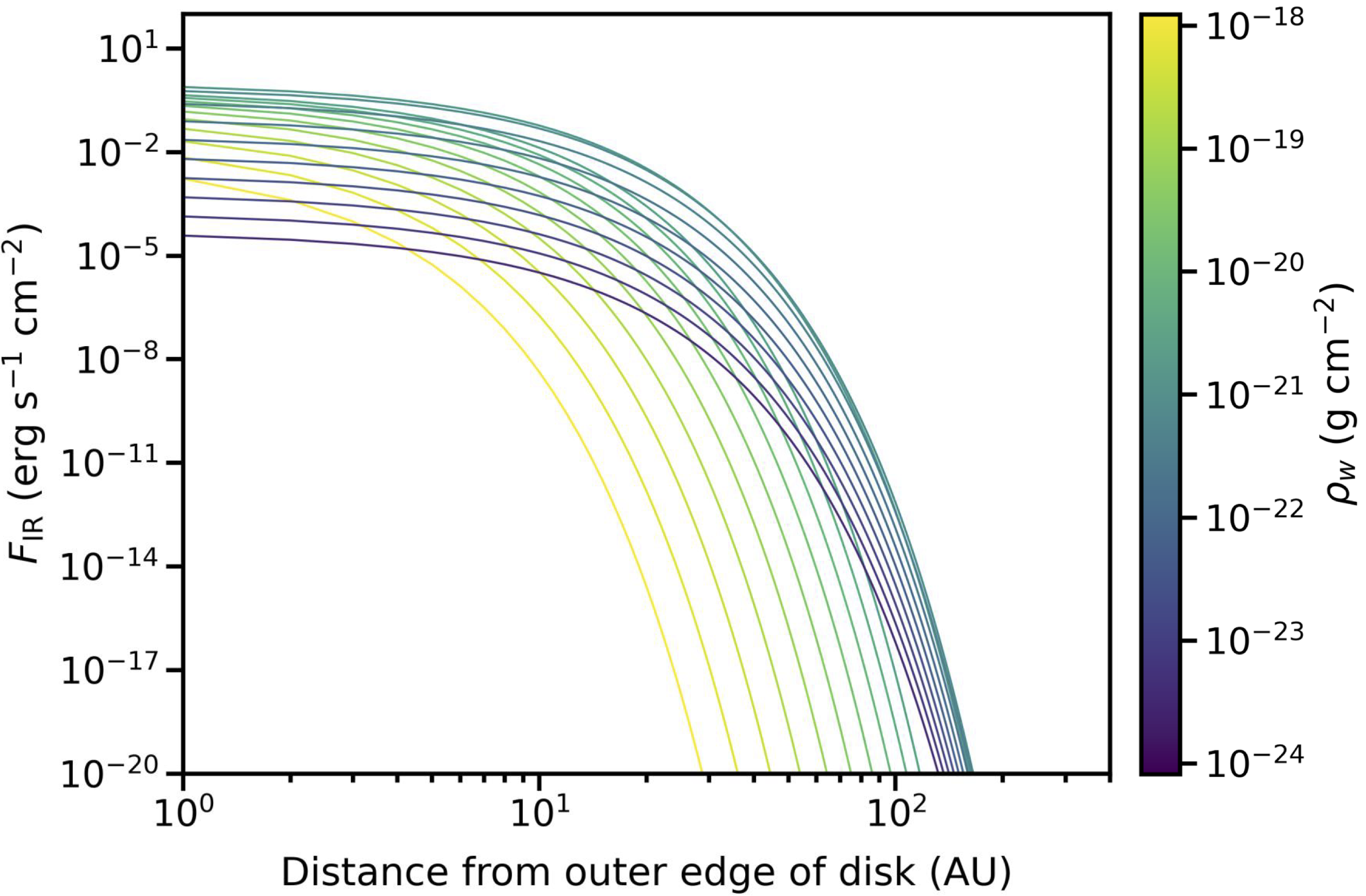}
\caption{Results of the 1D analytical model. Solid lines show the IR flux in the disk for wind dust densities ranging from $\rho_w=10^{-24}$ to $10^{-18}$ g cm$^{-3}$. Initially, increasing the dust density in the wind from the baseline value leads to enhanced IR intensity in the disk. However, at a critical turnover density ($\rho_\text{dust} \sim 10^{-21}$ g cm$^{-3}$), the IR emission from the wind is counterbalanced by increased absorption due to higher densities. This leads to two effects: weaker IR intensity and reduced penetration depth into the disk.}
\label{fig_1dRT_calc}
\end{figure}

We apply the model by setting the external UV field to 5000 G$_0$ ($\sim 1.35 \times 10^2$ erg s$^{-1}$ cm$^{-2}$), and adopt representative values for the UV and IR dust opacities. Based on typical values for small grain opacities at wavelengths of 912 \AA\ and 15 $\mu$m, we set $\kappa_\text{uv}=7.8 \times 10^4$ cm$^{2}$ g$^{-1}$ and $\kappa_\text{ir}=7.3 \times 10^2$ cm$^{2}$ g$^{-1}$ \citep{Bruderer2012}. The dust density in the disk region is set to $\rho_\text{dust} = 10^{-17}$ g cm$^{-3}$, derived from midplane values in our \textsc{dali} models. To investigate the impact of wind density on IR intensity within the disk, we systematically vary the wind dust density from $\rho_\text{dust} = 10^{-25}$ to $10^{-18}$ g cm$^{-3}$. The results of these calculations are illustrated in Figure \ref{fig_1dRT_calc}.

Our analysis reveals a nuanced relationship between wind dust density and IR intensity within the disk. As the dust density in the wind increases from the baseline value, an initial increase in IR intensity within the disk is observed, with radiation penetrating deeper into the disk. However, at a critical density threshold (approximately $\rho_\text{dust} \sim 10^{-21}$ g cm$^{-3}$), IR absorption by the wind becomes the dominant factor. Beyond this point, the IR radiation penetrating the disk diminishes, and its attenuation within the disk occurs more rapidly.

These results demonstrate that the presence of a photoevaporative wind can indeed lead to enhanced heating within the disk, but this effect is highly dependent on the wind's dust density. At low wind densities, the wind acts primarily as a reprocessing medium, absorbing UV photons and re-emitting them as IR radiation that can more effectively heat the disk. However, as the wind density increases beyond the critical threshold, it begins to act as a barrier, absorbing a significant portion of both UV and IR radiation before it can reach the disk. This highlights the need for future observational studies that can place constraints on key photoevaporative wind properties such as the dust-to-gas ratio and grain size distribution.

\subsection{Impact of model parameters}
\label{subsec:impact_of_params}

The chemical models presented in this work are based on outputs from complex hydrodynamical simulations, which are computationally expensive. Our exploration of the parameter space is therefore fundamentally limited by computational feasibility. In this section, we discuss how these limitations impact our results, and assess the significance of the specific parameter choices adopted for our models.

A key metric influencing disk evolution is its size, which is particularly significant for externally irradiated disks since the photoevaporative mass loss rate scales strongly with disk radius \citep[e.g.][]{haworth_2023_fried2}. Although a thorough exploration of the parameter space is beyond the scope of this work due to computational constraints, we conducted an additional simulation with an initial disk outer boundary set to $r_\text{d}=50$ au and an external UV field of 5000 G$_0$. While extended disks are commonly observed at the intermediate UV fields considered in this study (e.g., up to $\sim300$ au; \citealt{diaz_berrios_2024}), many externally irradiated disks tend to be more compact \citep[e.g.][]{mann_2014, eisner_2016, eisner_2018, ballering_2023, van_terwisga_2023}.
 
Figure \ref{fig_50au_disk_model} illustrates outputs from this $r_\text{d}=50$ au model, comparing cases where the wind is included and where it has been artificially removed. The results largely parallel those observed in the $r_\text{d}=100$ au configuration. When the wind is present, the $r_\text{d}=50$ au model exhibits temperature enhancements of several Kelvin, comparable to those observed in the larger disk model. This temperature difference is reflected in the gas-phase CO abundances, which are slightly elevated in the wind model, driven by increased thermal desorption. Consistent with the $r=100$ au model, UV radiation is rapidly attenuated by the disk in both the wind and no-wind scenarios for the $r_\text{d}=50$ au case. Analysis of synthetic line fluxes yields results analogous to those detailed in Section \ref{subsec:synthetic_obs}. The $r_\text{d}=50$ model demonstrates enhanced line fluxes when the wind is incorporated, with the magnitude of enhancement comparable to that observed in the $r_\text{d}=100$ model (approximately a factor of a few). The only exceptions are high-$J$ CO rotational transitions ($J \gtrsim 29)$, where the magnitude of flux increase becomes at least 100 times smaller than observed in the $r_\text{d}=100$ au model. 

Our results thus far have focused on FUV radiation, but EUV radiation could also play an important role in highly irradiated environments. While FUV penetrates deeper into the disk than EUV, the latter would establish an ionization front within the photoevaporative wind. In extreme cases where EUV radiation is sufficiently intense to ionize the wind down to our $n_\text{gas} = 10^8$ cm$^{-3}$ surface, the resulting scenario might resemble our `no wind' case, as an EUV-driven wind tends to be substantially more rarefied than its FUV counterpart. However, this would primarily affect the wind structure rather than the underlying disk chemistry, since both forms of radiation are efficiently attenuated before reaching the disk proper.

Stellar parameters represent another important consideration not fully explored in our current models. The properties of the central star influence both the photoevaporative wind structure (through its gravitational potential) and the competition between internal and external radiation fields in setting the disk chemistry. Higher mass stars would generate stronger winds and potentially alter the balance between internal heating and external irradiation, though we expect the general pattern of rapid UV attenuation near the disk surface to persist. Understanding this interplay between internal and external radiation fields is particularly important, as recent work by \citet{ramirez_tannus_2023} has highlighted challenges in distinguishing whether certain inner disk chemical signatures arise from domination by internal stellar radiation or from absorption of external FUV radiation by intervening material. Future chemical modelling studies that systematically explore stellar parameters could help disentangle these effects.

Our models consider UV fields up to 5000 G$_0$, but some systems experience substantially stronger irradiation. For instance, some of the most highly irradiated proplyds in the Orion Nebular Cluster are though to be exposed to FUV fields $G_0 \gtrsim 10^7$ \citep[e.g.][]{aru_2024}. At such extreme values, we anticipate that while the basic mechanisms identified in our study would still operate, the primary effect would be accelerated disk truncation and mass loss. The rapid attenuation of FUV radiation near the disk surface would likely continue to prevent direct UV heating from dominating the temperature structure, maintaining the importance of wind-driven IR heating. However, the more intense photoevaporation would lead to faster disk evolution and potentially different chemical outcomes in the rapidly depleting outer regions.

Finally, we also explored the sensitivity of our results to the number density that we use to demarcate the transition from `disk' to `wind' by systematically varying the disk-to-wind transition threshold ($n_\text{DWT}$) between $10^6$ - $10^9$ cm$^{-3}$. In each case we compared full-wind and no-wind scenarios. Our analysis reveals that, regardless of the exact position of the disk-wind boundary, the temperature variations between the full-wind and no-wind models persist but remain relatively small. However, the location of the boundary significantly influences the synthetic line fluxes. At higher transition densities ($n_\text{DWT} > 10^8$ cm$^{-3}$), the disk-wind interface shifts closer to the host star, resulting in a larger portion of the model grid being classified as wind. This leads to a more pronounced difference in line fluxes between the full-wind and no-wind scenarios. Conversely, at lower densities, the disk-wind boundary extends outward, reducing the line flux variations between the full-wind and no-wind models. These variations converge to the same value when $n_\text{DWT} \lesssim 10^6$ cm$^{-3}$, which corresponds to a radius of $\sim 500$ au at the midplane. Such a large disk radius is highly unrealistic when compared to the typical sizes of proplyds observed in the ONC.

Overall, these findings indicate that our results are therefore relatively robust to variations in key model parameters. However, a number of other important disk and stellar properties, such as disk mass, initial chemical composition, and stellar spectral type, remain unexplored. This highlights the need for parametric models of externally irradiated disks which include the wind component, facilitating a wider exploration of the parameter space that does not rely upon computationally expensive hydrodynamical simulations. Such parametric models are the focus of future work (in prep.).

\section{Conclusions}

In this work, we post-processed hydrodynamical simulations of externally irradiated protoplanetary disks with the \textsc{dali} thermochemical code to assess the impact of the photoevaporative wind on the disk chemistry. Our main findings are as follows.
\begin{enumerate}[leftmargin=*, align=left]
    \item The inclusion of a photoevaporative wind in the model's physical structure slightly enhances disk heating compared to windless models. This enhancement is not due to direct UV irradiation, as both wind and windless models rapidly attenuate the external UV field. Rather, the heating results from the wind's re-emission of absorbed UV radiation at infrared wavelengths, which penetrate deeper into the disk due to lower opacity.
    \item Gas-phase abundances of key volatiles (e.g., CO, CH$_4$, CN) are enhanced in wind-inclusive models, primarily due to increased thermal desorption rates. The location of these enhancements is species-dependent, with CO showing higher abundances near the midplane, while CH$_4$ exhibits an increase in the disk atmosphere.
    \item Contrary to other molecular species, the gas-phase H$_2$O abundance is enhanced in windless models. This enhancement is confined to a narrow vertical region in the disk atmosphere where the UV field is marginally higher than in wind models, leading to increased non-thermal desorption rates.
    \item Simulated disk-integrated line fluxes at infrared and millimetre wavelengths are typically greater in wind-inclusive models. Flux variations are species- and transition-dependent, ranging from factors of a few to over $10^5$. These higher fluxes primarily result from wind emission rather than abundance enhancements within the disk.
    \item The disparity in disk-integrated line flux between wind and windless models generally increases with higher rotational transitions. This indicates that higher $J$ transitions are more effective tracers of hot gas within the wind, providing a baseline for model-observation comparisons.
\end{enumerate}
Incorporating the photoevaporative wind into chemical models of externally irradiated disks is therefore important for accurate interpretation of observational data and prediction of chemical composition in these extreme planet-forming environments. Our findings underscore the need for 2D parametric models of externally irradiated disks that do not rely on computationally expensive hydrodynamic simulations, which we will deliver in a subsequent paper.


\section*{Acknowledgements}
We thank Catherine Walsh for useful discussions. We thank the reviewer for their helpful and insightful comments. LK is funded by UKRI guaranteed funding for a Horizon Europe ERC consolidator grant (EP/Y024710/1). TJH acknowledges UKRI guaranteed funding for a Horizon Europe ERC consolidator grant (EP/Y024710/1) and a Royal Society Dorothy Hodgkin Fellowship.

\section*{Data Availability}

Outputs from our chemical models are available on reasonable request from the corresponding author.


\bibliographystyle{mnras}
\bibliography{bibliography} 

\begin{thebibliography}{}
\makeatletter
\relax
\def\mn@urlcharsother{\let\do\@makeother \do\$\do\&\do\#\do\^\do\_\do\%\do\~}
\def\mn@doi{\begingroup\mn@urlcharsother \@ifnextchar [ {\mn@doi@} {\mn@doi@[]}}
\def\mn@doi@[#1]#2{\def\@tempa{#1}\ifx\@tempa\@empty \href {http://dx.doi.org/#2} {doi:#2}\else \href {http://dx.doi.org/#2} {#1}\fi \endgroup}
\def\mn@eprint#1#2{\mn@eprint@#1:#2::\@nil}
\def\mn@eprint@arXiv#1{\href {http://arxiv.org/abs/#1} {{\tt arXiv:#1}}}
\def\mn@eprint@dblp#1{\href {http://dblp.uni-trier.de/rec/bibtex/#1.xml} {dblp:#1}}
\def\mn@eprint@#1:#2:#3:#4\@nil{\def\@tempa {#1}\def\@tempb {#2}\def\@tempc {#3}\ifx \@tempc \@empty \let \@tempc \@tempb \let \@tempb \@tempa \fi \ifx \@tempb \@empty \def\@tempb {arXiv}\fi \@ifundefined {mn@eprint@\@tempb}{\@tempb:\@tempc}{\expandafter \expandafter \csname mn@eprint@\@tempb\endcsname \expandafter{\@tempc}}}

\bibitem[\protect\citeauthoryear{{Adams}, {Hollenbach}, {Laughlin}  \& {Gorti}}{{Adams} et~al.}{2004}]{adams_2004}
{Adams} F.~C.,  {Hollenbach} D.,  {Laughlin} G.,   {Gorti} U.,  2004, \mn@doi [\apj] {10.1086/421989}, \href {https://ui.adsabs.harvard.edu/abs/2004ApJ...611..360A} {611, 360}

\bibitem[\protect\citeauthoryear{{Andrews} et~al.,}{{Andrews} et~al.}{2018}]{DSHARPOverview}
{Andrews} S.~M.,  et~al., 2018, \mn@doi [\apjl] {10.3847/2041-8213/aaf741}, \href {https://ui.adsabs.harvard.edu/abs/2018ApJ...869L..41A} {869, L41}

\bibitem[\protect\citeauthoryear{{Ansdell} et~al.,}{{Ansdell} et~al.}{2016}]{ansdell_2016}
{Ansdell} M.,  et~al., 2016, \mn@doi [\apj] {10.3847/0004-637X/828/1/46}, \href {https://ui.adsabs.harvard.edu/abs/2016ApJ...828...46A} {828, 46}

\bibitem[\protect\citeauthoryear{{Aru} et~al.,}{{Aru} et~al.}{2024a}]{2024A&A...687A..93A}
{Aru} M.~L.,  et~al., 2024a, \mn@doi [\aap] {10.1051/0004-6361/202349004}, \href {https://ui.adsabs.harvard.edu/abs/2024A&A...687A..93A} {687, A93}

\bibitem[\protect\citeauthoryear{{Aru} et~al.,}{{Aru} et~al.}{2024b}]{aru_2024}
{Aru} M.~L.,  et~al., 2024b, \mn@doi [\aap] {10.1051/0004-6361/202451737}, \href {https://ui.adsabs.harvard.edu/abs/2024A&A...692A.137A} {692, A137}

\bibitem[\protect\citeauthoryear{{Ballabio}, {Haworth}  \& {Henney}}{{Ballabio} et~al.}{2023}]{ballabio_2023}
{Ballabio} G.,  {Haworth} T.~J.,   {Henney} W.~J.,  2023, \mn@doi [\mnras] {10.1093/mnras/stac3467}, \href {https://ui.adsabs.harvard.edu/abs/2023MNRAS.518.5563B} {518, 5563}

\bibitem[\protect\citeauthoryear{{Ballering}, {Cleeves}  \& {Anderson}}{{Ballering} et~al.}{2021}]{ballering_2021}
{Ballering} N.~P.,  {Cleeves} L.~I.,   {Anderson} D.~E.,  2021, \mn@doi [\apj] {10.3847/1538-4357/ac17ed}, \href {https://ui.adsabs.harvard.edu/abs/2021ApJ...920..115B} {920, 115}

\bibitem[\protect\citeauthoryear{{Ballering} et~al.,}{{Ballering} et~al.}{2023}]{ballering_2023}
{Ballering} N.~P.,  et~al., 2023, \mn@doi [\apj] {10.3847/1538-4357/ace901}, \href {https://ui.adsabs.harvard.edu/abs/2023ApJ...954..127B} {954, 127}

\bibitem[\protect\citeauthoryear{{Bally}, {Sutherland}, {Devine}  \& {Johnstone}}{{Bally} et~al.}{1998a}]{bally_1998a}
{Bally} J.,  {Sutherland} R.~S.,  {Devine} D.,   {Johnstone} D.,  1998a, \mn@doi [\aj] {10.1086/300399}, \href {https://ui.adsabs.harvard.edu/abs/1998AJ....116..293B} {116, 293}

\bibitem[\protect\citeauthoryear{{Bally}, {Testi}, {Sargent}  \& {Carlstrom}}{{Bally} et~al.}{1998b}]{bally_1998b}
{Bally} J.,  {Testi} L.,  {Sargent} A.,   {Carlstrom} J.,  1998b, \mn@doi [\aj] {10.1086/300469}, \href {https://ui.adsabs.harvard.edu/abs/1998AJ....116..854B} {116, 854}

\bibitem[\protect\citeauthoryear{{Bally}, {O'Dell}  \& {McCaughrean}}{{Bally} et~al.}{2000}]{bally_2000}
{Bally} J.,  {O'Dell} C.~R.,   {McCaughrean} M.~J.,  2000, \mn@doi [\aj] {10.1086/301385}, \href {https://ui.adsabs.harvard.edu/abs/2000AJ....119.2919B} {119, 2919}

\bibitem[\protect\citeauthoryear{{Bisbas}, {Bell}, {Viti}, {Yates}  \& {Barlow}}{{Bisbas} et~al.}{2012}]{bisbas_2012}
{Bisbas} T.~G.,  {Bell} T.~A.,  {Viti} S.,  {Yates} J.,   {Barlow} M.~J.,  2012, \mn@doi [\mnras] {10.1111/j.1365-2966.2012.22077.x}, \href {https://ui.adsabs.harvard.edu/abs/2012MNRAS.427.2100B} {427, 2100}

\bibitem[\protect\citeauthoryear{{Bisbas}, {Haworth}, {Barlow}, {Viti}, {Harries}, {Bell}  \& {Yates}}{{Bisbas} et~al.}{2015}]{bisbas_2015}
{Bisbas} T.~G.,  {Haworth} T.~J.,  {Barlow} M.~J.,  {Viti} S.,  {Harries} T.~J.,  {Bell} T.,   {Yates} J.~A.,  2015, \mn@doi [\mnras] {10.1093/mnras/stv2156}, \href {https://ui.adsabs.harvard.edu/abs/2015MNRAS.454.2828B} {454, 2828}

\bibitem[\protect\citeauthoryear{{Boehler} et~al.,}{{Boehler} et~al.}{2018}]{boehler_2018}
{Boehler} Y.,  et~al., 2018, \mn@doi [\apj] {10.3847/1538-4357/aaa19c}, \href {https://ui.adsabs.harvard.edu/abs/2018ApJ...853..162B} {853, 162}

\bibitem[\protect\citeauthoryear{{Booth} et~al.,}{{Booth} et~al.}{2024a}]{booth_2024a}
{Booth} A.~S.,  et~al., 2024a, \mn@doi [\aj] {10.3847/1538-3881/ad2700}, \href {https://ui.adsabs.harvard.edu/abs/2024AJ....167..164B} {167, 164}

\bibitem[\protect\citeauthoryear{{Booth} et~al.,}{{Booth} et~al.}{2024b}]{booth_2024b}
{Booth} A.~S.,  et~al., 2024b, \mn@doi [\aj] {10.3847/1538-3881/ad26ff}, \href {https://ui.adsabs.harvard.edu/abs/2024AJ....167..165B} {167, 165}

\bibitem[\protect\citeauthoryear{{Bosman} et~al.,}{{Bosman} et~al.}{2021}]{bosman_2021_maps}
{Bosman} A.~D.,  et~al., 2021, \mn@doi [\apjs] {10.3847/1538-4365/ac1435}, \href {https://ui.adsabs.harvard.edu/abs/2021ApJS..257....7B} {257, 7}

\bibitem[\protect\citeauthoryear{{Boyden} \& {Eisner}}{{Boyden} \& {Eisner}}{2020}]{boyden_eisner_2020}
{Boyden} R.~D.,  {Eisner} J.~A.,  2020, \mn@doi [\apj] {10.3847/1538-4357/ab86b7}, \href {https://ui.adsabs.harvard.edu/abs/2020ApJ...894...74B} {894, 74}

\bibitem[\protect\citeauthoryear{{Boyden} \& {Eisner}}{{Boyden} \& {Eisner}}{2023}]{boyden_eisner_2023}
{Boyden} R.~D.,  {Eisner} J.~A.,  2023, \mn@doi [\apj] {10.3847/1538-4357/acaf77}, \href {https://ui.adsabs.harvard.edu/abs/2023ApJ...947....7B} {947, 7}

\bibitem[\protect\citeauthoryear{{Bruderer}}{{Bruderer}}{2013}]{Bruderer2013}
{Bruderer} S.,  2013, \mn@doi [\aap] {10.1051/0004-6361/201321171}, \href {https://ui.adsabs.harvard.edu/abs/2013A&A...559A..46B} {559, A46}

\bibitem[\protect\citeauthoryear{{Bruderer}, {van Dishoeck}, {Doty}  \& {Herczeg}}{{Bruderer} et~al.}{2012}]{Bruderer2012}
{Bruderer} S.,  {van Dishoeck} E.~F.,  {Doty} S.~D.,   {Herczeg} G.~J.,  2012, \mn@doi [\aap] {10.1051/0004-6361/201118218}, \href {https://ui.adsabs.harvard.edu/abs/2012A&A...541A..91B} {541, A91}

\bibitem[\protect\citeauthoryear{{Chen}, {Bally}, {O'Dell}, {McCaughrean}, {Thompson}, {Rieke}, {Schneider}  \& {Young}}{{Chen} et~al.}{1998}]{chen_young_1998}
{Chen} H.,  {Bally} J.,  {O'Dell} C.~R.,  {McCaughrean} M.~J.,  {Thompson} R.~L.,  {Rieke} M.,  {Schneider} G.,   {Young} E.~T.,  1998, \mn@doi [\apjl] {10.1086/311098}, \href {https://ui.adsabs.harvard.edu/abs/1998ApJ...492L.173C} {492, L173}

\bibitem[\protect\citeauthoryear{{Clarke}}{{Clarke}}{2007}]{clarke_2007}
{Clarke} C.~J.,  2007, \mn@doi [\mnras] {10.1111/j.1365-2966.2007.11547.x}, \href {https://ui.adsabs.harvard.edu/abs/2007MNRAS.376.1350C} {376, 1350}

\bibitem[\protect\citeauthoryear{{Coleman} \& {Haworth}}{{Coleman} \& {Haworth}}{2022}]{ColemanAndHaworth22}
{Coleman} G. A.~L.,  {Haworth} T.~J.,  2022, \mn@doi [\mnras] {10.1093/mnras/stac1513}, \href {https://ui.adsabs.harvard.edu/abs/2022MNRAS.514.2315C} {514, 2315}

\bibitem[\protect\citeauthoryear{{Coleman}, {Mroueh}  \& {Haworth}}{{Coleman} et~al.}{2024}]{coleman_2024}
{Coleman} G. A.~L.,  {Mroueh} J.~K.,   {Haworth} T.~J.,  2024, \mn@doi [\mnras] {10.1093/mnras/stad3692}, \href {https://ui.adsabs.harvard.edu/abs/2024MNRAS.527.7588C} {527, 7588}

\bibitem[\protect\citeauthoryear{{D{\'\i}az-Berr{\'\i}os}, {Guzm{\'a}n}, {Walsh}, {{\"O}berg}, {Cleeves}, {Artur de la Villarmois}  \& {Carpenter}}{{D{\'\i}az-Berr{\'\i}os} et~al.}{2024}]{diaz_berrios_2024}
{D{\'\i}az-Berr{\'\i}os} J.~K.,  {Guzm{\'a}n} V.~V.,  {Walsh} C.,  {{\"O}berg} K.~I.,  {Cleeves} L.~I.,  {Artur de la Villarmois} E.,   {Carpenter} J.,  2024, \mn@doi [\apj] {10.3847/1538-4357/ad4603}, \href {https://ui.adsabs.harvard.edu/abs/2024ApJ...969..165D} {969, 165}

\bibitem[\protect\citeauthoryear{{Eisner} \& {Carpenter}}{{Eisner} \& {Carpenter}}{2006}]{eisner_2006}
{Eisner} J.~A.,  {Carpenter} J.~M.,  2006, \mn@doi [\apj] {10.1086/500637}, \href {https://ui.adsabs.harvard.edu/abs/2006ApJ...641.1162E} {641, 1162}

\bibitem[\protect\citeauthoryear{{Eisner}, {Bally}, {Ginsburg}  \& {Sheehan}}{{Eisner} et~al.}{2016}]{eisner_2016}
{Eisner} J.~A.,  {Bally} J.~M.,  {Ginsburg} A.,   {Sheehan} P.~D.,  2016, \mn@doi [\apj] {10.3847/0004-637X/826/1/16}, \href {https://ui.adsabs.harvard.edu/abs/2016ApJ...826...16E} {826, 16}

\bibitem[\protect\citeauthoryear{{Eisner} et~al.,}{{Eisner} et~al.}{2018}]{eisner_2018}
{Eisner} J.~A.,  et~al., 2018, \mn@doi [\apj] {10.3847/1538-4357/aac3e2}, \href {https://ui.adsabs.harvard.edu/abs/2018ApJ...860...77E} {860, 77}

\bibitem[\protect\citeauthoryear{{Facchini}, {Clarke}  \& {Bisbas}}{{Facchini} et~al.}{2016}]{facchini_2016}
{Facchini} S.,  {Clarke} C.~J.,   {Bisbas} T.~G.,  2016, \mn@doi [\mnras] {10.1093/mnras/stw240}, \href {https://ui.adsabs.harvard.edu/abs/2016MNRAS.457.3593F} {457, 3593}

\bibitem[\protect\citeauthoryear{{Fatuzzo} \& {Adams}}{{Fatuzzo} \& {Adams}}{2008}]{FatuzzoAndAdams2008}
{Fatuzzo} M.,  {Adams} F.~C.,  2008, \mn@doi [\apj] {10.1086/527469}, \href {https://ui.adsabs.harvard.edu/abs/2008ApJ...675.1361F} {675, 1361}

\bibitem[\protect\citeauthoryear{{Fedele} et~al.,}{{Fedele} et~al.}{2017}]{fedele_2017}
{Fedele} D.,  et~al., 2017, \mn@doi [\aap] {10.1051/0004-6361/201629860}, \href {https://ui.adsabs.harvard.edu/abs/2017A&A...600A..72F} {600, A72}

\bibitem[\protect\citeauthoryear{{Goicoechea} et~al.,}{{Goicoechea} et~al.}{2024}]{goicoechea_2024}
{Goicoechea} J.~R.,  et~al., 2024, \mn@doi [\aap] {10.1051/0004-6361/202450988}, \href {https://ui.adsabs.harvard.edu/abs/2024A&A...689L...4G} {689, L4}

\bibitem[\protect\citeauthoryear{{Haworth} \& {Clarke}}{{Haworth} \& {Clarke}}{2019}]{haworth_clarke_2019}
{Haworth} T.~J.,  {Clarke} C.~J.,  2019, \mn@doi [\mnras] {10.1093/mnras/stz706}, \href {https://ui.adsabs.harvard.edu/abs/2019MNRAS.485.3895H} {485, 3895}

\bibitem[\protect\citeauthoryear{{Haworth} \& {Harries}}{{Haworth} \& {Harries}}{2012}]{haworth_harries_2012}
{Haworth} T.~J.,  {Harries} T.~J.,  2012, \mn@doi [\mnras] {10.1111/j.1365-2966.2011.20062.x}, \href {https://ui.adsabs.harvard.edu/abs/2012MNRAS.420..562H} {420, 562}

\bibitem[\protect\citeauthoryear{{Haworth}, {Clarke}, {Rahman}, {Winter}  \& {Facchini}}{{Haworth} et~al.}{2018}]{haworth_2018_fried}
{Haworth} T.~J.,  {Clarke} C.~J.,  {Rahman} W.,  {Winter} A.~J.,   {Facchini} S.,  2018, \mn@doi [\mnras] {10.1093/mnras/sty2323}, \href {https://ui.adsabs.harvard.edu/abs/2018MNRAS.481..452H} {481, 452}

\bibitem[\protect\citeauthoryear{{Haworth}, {Kim}, {Winter}, {Hines}, {Clarke}, {Sellek}, {Ballabio}  \& {Stapelfeldt}}{{Haworth} et~al.}{2021}]{2021MNRAS.501.3502H}
{Haworth} T.~J.,  {Kim} J.~S.,  {Winter} A.~J.,  {Hines} D.~C.,  {Clarke} C.~J.,  {Sellek} A.~D.,  {Ballabio} G.,   {Stapelfeldt} K.~R.,  2021, \mn@doi [\mnras] {10.1093/mnras/staa3918}, \href {https://ui.adsabs.harvard.edu/abs/2021MNRAS.501.3502H} {501, 3502}

\bibitem[\protect\citeauthoryear{{Haworth}, {Coleman}, {Qiao}, {Sellek}  \& {Askari}}{{Haworth} et~al.}{2023}]{haworth_2023_fried2}
{Haworth} T.~J.,  {Coleman} G. A.~L.,  {Qiao} L.,  {Sellek} A.~D.,   {Askari} K.,  2023, \mn@doi [\mnras] {10.1093/mnras/stad3054}, \href {https://ui.adsabs.harvard.edu/abs/2023MNRAS.526.4315H} {526, 4315}

\bibitem[\protect\citeauthoryear{{Henney} \& {O'Dell}}{{Henney} \& {O'Dell}}{1999}]{Henney_1999}
{Henney} W.~J.,  {O'Dell} C.~R.,  1999, \mn@doi [\aj] {10.1086/301087}, \href {https://ui.adsabs.harvard.edu/abs/1999AJ....118.2350H} {118, 2350}

\bibitem[\protect\citeauthoryear{{Hillenbrand}, {Strom}, {Calvet}, {Merrill}, {Gatley}, {Makidon}, {Meyer}  \& {Skrutskie}}{{Hillenbrand} et~al.}{1998}]{hillenbrand_1998}
{Hillenbrand} L.~A.,  {Strom} S.~E.,  {Calvet} N.,  {Merrill} K.~M.,  {Gatley} I.,  {Makidon} R.~B.,  {Meyer} M.~R.,   {Skrutskie} M.~F.,  1998, \mn@doi [\aj] {10.1086/300536}, \href {https://ui.adsabs.harvard.edu/abs/1998AJ....116.1816H} {116, 1816}

\bibitem[\protect\citeauthoryear{{Huang}, {Portegies Zwart}  \& {Wilhelm}}{{Huang} et~al.}{2024}]{HuangPF}
{Huang} S.,  {Portegies Zwart} S.,   {Wilhelm} M. J.~C.,  2024, \mn@doi [\aap] {10.1051/0004-6361/202451051}, \href {https://ui.adsabs.harvard.edu/abs/2024A&A...689A.338H} {689, A338}

\bibitem[\protect\citeauthoryear{{Isella} et~al.,}{{Isella} et~al.}{2016}]{isella_2016}
{Isella} A.,  et~al., 2016, \mn@doi [\prl] {10.1103/PhysRevLett.117.251101}, \href {https://ui.adsabs.harvard.edu/abs/2016PhRvL.117y1101I} {117, 251101}

\bibitem[\protect\citeauthoryear{{Johnstone}, {Hollenbach}  \& {Bally}}{{Johnstone} et~al.}{1998}]{johnstone_1998}
{Johnstone} D.,  {Hollenbach} D.,   {Bally} J.,  1998, \mn@doi [\apj] {10.1086/305658}, \href {https://ui.adsabs.harvard.edu/abs/1998ApJ...499..758J} {499, 758}

\bibitem[\protect\citeauthoryear{{Kama} et~al.,}{{Kama} et~al.}{2016a}]{kama_2016a}
{Kama} M.,  et~al., 2016a, \mn@doi [\aap] {10.1051/0004-6361/201526791}, \href {https://ui.adsabs.harvard.edu/abs/2016A&A...588A.108K} {588, A108}

\bibitem[\protect\citeauthoryear{{Kama} et~al.,}{{Kama} et~al.}{2016b}]{kama_2016b}
{Kama} M.,  et~al., 2016b, \mn@doi [\aap] {10.1051/0004-6361/201526991}, \href {https://ui.adsabs.harvard.edu/abs/2016A&A...592A..83K} {592, A83}

\bibitem[\protect\citeauthoryear{{Kennicutt} \& {Evans}}{{Kennicutt} \& {Evans}}{2012}]{kennicutt_evans_2012}
{Kennicutt} R.~C.,  {Evans} N.~J.,  2012, \mn@doi [\araa] {10.1146/annurev-astro-081811-125610}, \href {https://ui.adsabs.harvard.edu/abs/2012ARA&A..50..531K} {50, 531}

\bibitem[\protect\citeauthoryear{{Keyte} et~al.,}{{Keyte} et~al.}{2023}]{keyte_2023}
{Keyte} L.,  et~al., 2023, \mn@doi [Nature Astronomy] {10.1038/s41550-023-01951-9}, \href {https://ui.adsabs.harvard.edu/abs/2023NatAs...7..684K} {7, 684}

\bibitem[\protect\citeauthoryear{{Keyte}, {Kama}, {Booth}, {Law}  \& {Leemker}}{{Keyte} et~al.}{2024a}]{keyte_2024b}
{Keyte} L.,  {Kama} M.,  {Booth} A.~S.,  {Law} C.~J.,   {Leemker} M.,  2024a, \mn@doi [\mnras] {10.1093/mnras/stae2314}, \href {https://ui.adsabs.harvard.edu/abs/2024MNRAS.tmp.2274K} {}

\bibitem[\protect\citeauthoryear{{Keyte}, {Kama}, {Chuang}, {Cleeves}, {Drozdovskaya}, {Furuya}, {Rawlings}  \& {Shorttle}}{{Keyte} et~al.}{2024b}]{keyte_2024a}
{Keyte} L.,  {Kama} M.,  {Chuang} K.-J.,  {Cleeves} L.~I.,  {Drozdovskaya} M.~N.,  {Furuya} K.,  {Rawlings} J.,   {Shorttle} O.,  2024b, \mn@doi [\mnras] {10.1093/mnras/stae019}, \href {https://ui.adsabs.harvard.edu/abs/2024MNRAS.528..388K} {528, 388}

\bibitem[\protect\citeauthoryear{{Kim}, {Clarke}, {Fang}  \& {Facchini}}{{Kim} et~al.}{2016}]{2016ApJ...826L..15K}
{Kim} J.~S.,  {Clarke} C.~J.,  {Fang} M.,   {Facchini} S.,  2016, \mn@doi [\apjl] {10.3847/2041-8205/826/1/L15}, \href {https://ui.adsabs.harvard.edu/abs/2016ApJ...826L..15K} {826, L15}

\bibitem[\protect\citeauthoryear{{Krumholz}, {McKee}  \& {Bland-Hawthorn}}{{Krumholz} et~al.}{2019}]{krumholz_2019}
{Krumholz} M.~R.,  {McKee} C.~F.,   {Bland-Hawthorn} J.,  2019, \mn@doi [\araa] {10.1146/annurev-astro-091918-104430}, \href {https://ui.adsabs.harvard.edu/abs/2019ARA&A..57..227K} {57, 227}

\bibitem[\protect\citeauthoryear{{Lada} \& {Lada}}{{Lada} \& {Lada}}{2003}]{lada_lada_2003}
{Lada} C.~J.,  {Lada} E.~A.,  2003, \mn@doi [\araa] {10.1146/annurev.astro.41.011802.094844}, \href {https://ui.adsabs.harvard.edu/abs/2003ARA&A..41...57L} {41, 57}

\bibitem[\protect\citeauthoryear{{Lazareff} et~al.,}{{Lazareff} et~al.}{2017}]{lazareff_2017}
{Lazareff} B.,  et~al., 2017, \mn@doi [\aap] {10.1051/0004-6361/201629305}, \href {https://ui.adsabs.harvard.edu/abs/2017A&A...599A..85L} {599, A85}

\bibitem[\protect\citeauthoryear{{Leemker} et~al.,}{{Leemker} et~al.}{2022}]{leemker_2022}
{Leemker} M.,  et~al., 2022, \mn@doi [\aap] {10.1051/0004-6361/202243229}, \href {https://ui.adsabs.harvard.edu/abs/2022A&A...663A..23L} {663, A23}

\bibitem[\protect\citeauthoryear{{Long} et~al.,}{{Long} et~al.}{2018}]{long_2018}
{Long} F.,  et~al., 2018, \mn@doi [\apj] {10.3847/1538-4357/aae8e1}, \href {https://ui.adsabs.harvard.edu/abs/2018ApJ...869...17L} {869, 17}

\bibitem[\protect\citeauthoryear{{Lynden-Bell} \& {Pringle}}{{Lynden-Bell} \& {Pringle}}{1974}]{lyndenbell_pringle_1974}
{Lynden-Bell} D.,  {Pringle} J.~E.,  1974, \mn@doi [\mnras] {10.1093/mnras/168.3.603}, \href {https://ui.adsabs.harvard.edu/abs/1974MNRAS.168..603L} {168, 603}

\bibitem[\protect\citeauthoryear{{Mann} et~al.,}{{Mann} et~al.}{2014}]{mann_2014}
{Mann} R.~K.,  et~al., 2014, \mn@doi [\apj] {10.1088/0004-637X/784/1/82}, \href {https://ui.adsabs.harvard.edu/abs/2014ApJ...784...82M} {784, 82}

\bibitem[\protect\citeauthoryear{{Mathis}, {Rumpl}  \& {Nordsieck}}{{Mathis} et~al.}{1977}]{mathis_1977}
{Mathis} J.~S.,  {Rumpl} W.,   {Nordsieck} K.~H.,  1977, \mn@doi [\apj] {10.1086/155591}, \href {https://ui.adsabs.harvard.edu/abs/1977ApJ...217..425M} {217, 425}

\bibitem[\protect\citeauthoryear{{McCaughrean} \& {O'Dell}}{{McCaughrean} \& {O'Dell}}{1996}]{mccaughrean_odell_1996}
{McCaughrean} M.~J.,  {O'Dell} C.~R.,  1996, \mn@doi [\aj] {10.1086/117934}, \href {https://ui.adsabs.harvard.edu/abs/1996AJ....111.1977M} {111, 1977}

\bibitem[\protect\citeauthoryear{{McElroy}, {Walsh}, {Markwick}, {Cordiner}, {Smith}  \& {Millar}}{{McElroy} et~al.}{2013}]{mcelroy_2013_umist}
{McElroy} D.,  {Walsh} C.,  {Markwick} A.~J.,  {Cordiner} M.~A.,  {Smith} K.,   {Millar} T.~J.,  2013, \mn@doi [\aap] {10.1051/0004-6361/201220465}, \href {https://ui.adsabs.harvard.edu/abs/2013A&A...550A..36M} {550, A36}

\bibitem[\protect\citeauthoryear{{Moullet} et~al.,}{{Moullet} et~al.}{2023}]{moullet_2023_prima}
{Moullet} A.,  et~al., 2023, \mn@doi [arXiv e-prints] {10.48550/arXiv.2310.20572}, \href {https://ui.adsabs.harvard.edu/abs/2023arXiv231020572M} {p. arXiv:2310.20572}

\bibitem[\protect\citeauthoryear{{O'Dell} \& {Wen}}{{O'Dell} \& {Wen}}{1994}]{odell_wen_1994}
{O'Dell} C.~R.,  {Wen} Z.,  1994, \mn@doi [\apj] {10.1086/174892}, \href {https://ui.adsabs.harvard.edu/abs/1994ApJ...436..194O} {436, 194}

\bibitem[\protect\citeauthoryear{{O'Dell}, {Wen}  \& {Hu}}{{O'Dell} et~al.}{1993}]{odell_wen_hu_1993}
{O'Dell} C.~R.,  {Wen} Z.,   {Hu} X.,  1993, \mn@doi [\apj] {10.1086/172786}, \href {https://ui.adsabs.harvard.edu/abs/1993ApJ...410..696O} {410, 696}

\bibitem[\protect\citeauthoryear{{{\"O}berg} et~al.,}{{{\"O}berg} et~al.}{2021}]{MAPSOverview}
{{\"O}berg} K.~I.,  et~al., 2021, \mn@doi [\apjs] {10.3847/1538-4365/ac1432}, \href {https://ui.adsabs.harvard.edu/abs/2021ApJS..257....1O} {257, 1}

\bibitem[\protect\citeauthoryear{{Owen} \& {Altaf}}{{Owen} \& {Altaf}}{2021}]{2021MNRAS.508.2493O}
{Owen} J.~E.,  {Altaf} N.,  2021, \mn@doi [\mnras] {10.1093/mnras/stab2749}, \href {https://ui.adsabs.harvard.edu/abs/2021MNRAS.508.2493O} {508, 2493}

\bibitem[\protect\citeauthoryear{{Pilbratt} et~al.,}{{Pilbratt} et~al.}{2010}]{pilbratt_2010_herschel}
{Pilbratt} G.~L.,  et~al., 2010, \mn@doi [\aap] {10.1051/0004-6361/201014759}, \href {https://ui.adsabs.harvard.edu/abs/2010A&A...518L...1P} {518, L1}

\bibitem[\protect\citeauthoryear{{Qiao}, {Coleman}  \& {Haworth}}{{Qiao} et~al.}{2023}]{QiaoPF}
{Qiao} L.,  {Coleman} G. A.~L.,   {Haworth} T.~J.,  2023, \mn@doi [\mnras] {10.1093/mnras/stad944}, \href {https://ui.adsabs.harvard.edu/abs/2023MNRAS.522.1939Q} {522, 1939}

\bibitem[\protect\citeauthoryear{{Ram{\'\i}rez-Tannus} et~al.,}{{Ram{\'\i}rez-Tannus} et~al.}{2023}]{ramirez_tannus_2023}
{Ram{\'\i}rez-Tannus} M.~C.,  et~al., 2023, \mn@doi [\apjl] {10.3847/2041-8213/ad03f8}, \href {https://ui.adsabs.harvard.edu/abs/2023ApJ...958L..30R} {958, L30}

\bibitem[\protect\citeauthoryear{{Ricci}, {Robberto}  \& {Soderblom}}{{Ricci} et~al.}{2008}]{Ricci_2008}
{Ricci} L.,  {Robberto} M.,   {Soderblom} D.~R.,  2008, \mn@doi [\aj] {10.1088/0004-6256/136/5/2136}, \href {https://ui.adsabs.harvard.edu/abs/2008AJ....136.2136R} {136, 2136}

\bibitem[\protect\citeauthoryear{{Sandford} \& {Allamandola}}{{Sandford} \& {Allamandola}}{1993}]{sandford_allamandola_1993}
{Sandford} S.~A.,  {Allamandola} L.~J.,  1993, \mn@doi [\apj] {10.1086/173362}, \href {https://ui.adsabs.harvard.edu/abs/1993ApJ...417..815S} {417, 815}

\bibitem[\protect\citeauthoryear{{Sch{\"o}ier}, {van der Tak}, {van Dishoeck}  \& {Black}}{{Sch{\"o}ier} et~al.}{2005}]{schoier_LAMDA_2005}
{Sch{\"o}ier} F.~L.,  {van der Tak} F.~F.~S.,  {van Dishoeck} E.~F.,   {Black} J.~H.,  2005, \mn@doi [\aap] {10.1051/0004-6361:20041729}, \href {https://ui.adsabs.harvard.edu/abs/2005A&A...432..369S} {432, 369}

\bibitem[\protect\citeauthoryear{{Stapper} et~al.,}{{Stapper} et~al.}{2024}]{stapper_2024}
{Stapper} L.~M.,  et~al., 2024, \mn@doi [\aap] {10.1051/0004-6361/202347271}, \href {https://ui.adsabs.harvard.edu/abs/2024A&A...682A.149S} {682, A149}

\bibitem[\protect\citeauthoryear{{Trapman}, {Rosotti}, {Bosman}, {Hogerheijde}  \& {van Dishoeck}}{{Trapman} et~al.}{2020}]{trapman_2020b}
{Trapman} L.,  {Rosotti} G.,  {Bosman} A.~D.,  {Hogerheijde} M.~R.,   {van Dishoeck} E.~F.,  2020, \mn@doi [\aap] {10.1051/0004-6361/202037673}, \href {https://ui.adsabs.harvard.edu/abs/2020A&A...640A...5T} {640, A5}

\bibitem[\protect\citeauthoryear{{Vicente} \& {Alves}}{{Vicente} \& {Alves}}{2005}]{vicente_alves_2005}
{Vicente} S.~M.,  {Alves} J.,  2005, \mn@doi [\aap] {10.1051/0004-6361:20053540}, \href {https://ui.adsabs.harvard.edu/abs/2005A&A...441..195V} {441, 195}

\bibitem[\protect\citeauthoryear{{Visser}, {Doty}  \& {van Dishoeck}}{{Visser} et~al.}{2011}]{visser_2011}
{Visser} R.,  {Doty} S.~D.,   {van Dishoeck} E.~F.,  2011, \mn@doi [\aap] {10.1051/0004-6361/201117249}, \href {https://ui.adsabs.harvard.edu/abs/2011A&A...534A.132V} {534, A132}

\bibitem[\protect\citeauthoryear{{Visser}, {Bruderer}, {Cazzoletti}, {Facchini}, {Heays}  \& {van Dishoeck}}{{Visser} et~al.}{2018}]{visser_2018}
{Visser} R.,  {Bruderer} S.,  {Cazzoletti} P.,  {Facchini} S.,  {Heays} A.~N.,   {van Dishoeck} E.~F.,  2018, \mn@doi [\aap] {10.1051/0004-6361/201731898}, \href {https://ui.adsabs.harvard.edu/abs/2018A&A...615A..75V} {615, A75}

\bibitem[\protect\citeauthoryear{{Walsh}, {Millar}  \& {Nomura}}{{Walsh} et~al.}{2013}]{walsh_2013}
{Walsh} C.,  {Millar} T.~J.,   {Nomura} H.,  2013, \mn@doi [\apjl] {10.1088/2041-8205/766/2/L23}, \href {https://ui.adsabs.harvard.edu/abs/2013ApJ...766L..23W} {766, L23}

\bibitem[\protect\citeauthoryear{{Winter} \& {Haworth}}{{Winter} \& {Haworth}}{2022}]{winter_haworth_2022}
{Winter} A.~J.,  {Haworth} T.~J.,  2022, \mn@doi [European Physical Journal Plus] {10.1140/epjp/s13360-022-03314-1}, \href {https://ui.adsabs.harvard.edu/abs/2022EPJP..137.1132W} {137, 1132}

\bibitem[\protect\citeauthoryear{{Winter}, {Kruijssen}, {Chevance}, {Keller}  \& {Longmore}}{{Winter} et~al.}{2020}]{WinterPrevalence}
{Winter} A.~J.,  {Kruijssen} J.~M.~D.,  {Chevance} M.,  {Keller} B.~W.,   {Longmore} S.~N.,  2020, \mn@doi [\mnras] {10.1093/mnras/stz2747}, \href {https://ui.adsabs.harvard.edu/abs/2020MNRAS.491..903W} {491, 903}

\bibitem[\protect\citeauthoryear{{Winter}, {Haworth}, {Coleman}  \& {Nayakshin}}{{Winter} et~al.}{2022}]{WinterPF}
{Winter} A.~J.,  {Haworth} T.~J.,  {Coleman} G. A.~L.,   {Nayakshin} S.,  2022, \mn@doi [\mnras] {10.1093/mnras/stac1564}, \href {https://ui.adsabs.harvard.edu/abs/2022MNRAS.515.4287W} {515, 4287}

\bibitem[\protect\citeauthoryear{{Woodall}, {Ag{\'u}ndez}, {Markwick-Kemper}  \& {Millar}}{{Woodall} et~al.}{2007}]{woodall2007}
{Woodall} J.,  {Ag{\'u}ndez} M.,  {Markwick-Kemper} A.~J.,   {Millar} T.~J.,  2007, \mn@doi [\aap] {10.1051/0004-6361:20064981}, \href {https://ui.adsabs.harvard.edu/abs/2007A&A...466.1197W} {466, 1197}

\bibitem[\protect\citeauthoryear{{van Terwisga} \& {Hacar}}{{van Terwisga} \& {Hacar}}{2023}]{van_terwisga_2023}
{van Terwisga} S.~E.,  {Hacar} A.,  2023, \mn@doi [\aap] {10.1051/0004-6361/202346135}, \href {https://ui.adsabs.harvard.edu/abs/2023A&A...673L...2V} {673, L2}

\bibitem[\protect\citeauthoryear{{van Terwisga}, {Hacar}  \& {van Dishoeck}}{{van Terwisga} et~al.}{2019}]{van_terwisga_2019}
{van Terwisga} S.~E.,  {Hacar} A.,   {van Dishoeck} E.~F.,  2019, \mn@doi [\aap] {10.1051/0004-6361/201935378}, \href {https://ui.adsabs.harvard.edu/abs/2019A&A...628A..85V} {628, A85}

\makeatother
\end{thebibliography}



\appendix

\section{DALI model parameters}
\label{apx:dali_model}

\begin{table*}
\caption{DALI model parameters.}             
\label{table:modelparameters}      
\centering
\begin{tabular}{l l l}     
\hline\hline       
                      
Parameter & Value & Description \\ 
\hline\hline      
\emph{Physical structure} & & \\
\hline
   $R_\text{in}$        & 1 au       & Inner disk boundary    \\
   $R_\text{out}$       & 800 au     & Outer grid boundary    \\
   $z_\text{max}$       & 850 au     & Upper grid boundary    \\
\hline
\emph{Dust and PAH properties} & & \\
\hline
   $\chi$                         & 0.2                   & Dust settling parameter                \\
   $f_0$                          & 0.9                   & Midplane large grain mass fraction     \\
   $\Delta_\text{d/g}$ (disk)     & $1 \times 10^{-2}$    & Dust-to-gas ratio in the disk          \\
   $\Delta_\text{d/g}$ (wind)     & $3 \times 10^{-4}$    & Dust-to-gas ratio in the wind          \\
   PAH abundance                  & 1.0                   & Abundance of PAHs relative to the ISM  \\
\hline
\emph{Stellar and radiation field properties} & & \\
\hline
   $M_*$                & $1\; M_\odot$                               & Stellar mass           \\
   $L_*$                & $1\; L_\odot$                               & Stellar luminosity           \\
   $T_*$                & $5500$ K                                    & Stellar temperature          \\
   $L_X$                & $7.94 \times 10^{28} \text{ erg s}^{-1}$    & Stellar X-ray luminosity     \\
   $T_X$                & $7.0 \times 10^{7}$ K                       & X-ray plasma temperature     \\
   $\zeta_\text{cr}$    & $1.26 \times 10^{-17}$ s$^{-1}$             & Cosmic ray ionization rate   \\
\hline
\emph{Chemistry} & & \\
\hline
   H$_2$                & $5.00 \times 10^{-1}$ & Initial H$_2$ abundance         \\
   He                   & $9.00 \times 10^{-2}$ & Initial He abundance            \\
   N$_2$                & $3.51 \times 10^{-5}$ & Initial N$_2$ abundance         \\
   H$_2$O               & $1.00 \times 10^{-4}$ & Initial H$_2$O abundance        \\
   CO                   & $1.00 \times 10^{-4}$ & Initial CO abundance            \\
   C$^+$                & $1.00 \times 10^{-5}$ & Initial C$^+$ abundance         \\
   NH$_3$               & $4.80 \times 10^{-6}$ & Initial NH$_3$ abundance        \\
   HCN                  & $2.00 \times 10^{-8}$ & Initial HCN abundance           \\
   H$_3^+$              & $1.00 \times 10^{-8}$ & Initial H$_3^+$ abundance       \\
   HCO$^+$              & $9.00 \times 10^{-9}$ & Initial HCO$^+$ abundance       \\
   C$_2$H               & $8.00 \times 10^{-9}$ & Initial C$_2$H abundances       \\
   S$^+$                & $9.00 \times 10^{-9}$ & Initial S$^+$ abundance         \\
   Si$^+$               & $1.00 \times 10^{-11}$ & Initial Si$^+$ abundance       \\
   Mg$^+$               & $1.00 \times 10^{-11}$ & Initial Mg$^+$ abundance       \\
   Fe$^+$               & $1.00 \times 10^{-11}$ & Initial Fe$^+$ abundance       \\
   $t_\text{chem}$      & \text{1 Myr} & Timescale for time-dependent chemistry   \\
\hline                  
\end{tabular}
\end{table*}

\clearpage

\section{Synthetic line fluxes and contribution functions}
\label{apx:linefluxes_cbfs}
\onecolumn
\begin{small}
\begin{longtable}{lllllll}
\caption{Molecular transitions and line fluxes incorporated in our modelling. Line transition data taken from the Leiden Atomic and Molecular Database \citep[LAMDA; ][]{schoier_LAMDA_2005}.} \label{table:line_fluxes} \\
\hline \hline 
Molecule & Transition & Frequency (GHz) & Wavelength ($\mu$m) & $F_\text{wind}$ (W m$^{-2}$) & $F_\text{no-wind}$ (W m$^{-2}$) & $F_\text{wind}/F_\text{no-wind}$ \\
\hline 
\endfirsthead

\hline \hline 
Molecule & Transition & Frequency (GHz) & Wavelength ($\mu$m) & $F_\text{wind}$ (W m$^{-2}$) & $F_\text{no-wind}$ (W m$^{-2}$) & $F_\text{wind}/F_\text{no-wind}$ \\
\hline 
\endhead

\hline 
\endfoot
$[$CI$]$ & $^3P_1-^3P_0$ & 492.161 & 609.14 & $1.98 \times 10^{-19}$ & $1.06 \times 10^{-22}$ & 1868.2 \\
$[$CI$]$ & $^3P_2-^3P_1$ & 809.342 & 370.42 & $1.59 \times 10^{-18}$ & $3.29 \times 10^{-22}$ & 4829.4 \\
C$_2$H & $N=1-0, \; J=3/2-1/2$ & 87.402 & 3430.04 & $4.21 \times 10^{-24}$ & $1.23 \times 10^{-25}$ & 34.4 \\
C$_2$H & $N=2-1, \; J=5/2-3/2$ & 174.663 & 1716.40 & $2.31 \times 10^{-22}$ & $4.85 \times 10^{-24}$ & 47.5 \\
C$_2$H & $N=3-2, \; J=7/2-5/2$ & 262.004 & 1143.23 & $1.27 \times 10^{-21}$ & $2.19 \times 10^{-23}$ & 57.8 \\
C$_2$H & $N=4-3, \; J=9/2-7/2$ & 349.338 & 858.17 & $3.04 \times 10^{-21}$ & $4.98 \times 10^{-23}$ & 61.1 \\
C$_2$H & $N=5-4, \; J=11/2-9/2$ & 436.661 & 686.56 & $4.48 \times 10^{-21}$ & $7.37 \times 10^{-23}$ & 60.7 \\
C$_2$H & $N=6-5, \; J=13/2-11/2$ & 523.972 & 572.15 & $4.52 \times 10^{-21}$ & $7.88 \times 10^{-23}$ & 57.4 \\
C$_2$H & $N=7-6, \; J=15/2-13/2$ & 611.267 & 490.44 & $3.50 \times 10^{-21}$ & $6.73 \times 10^{-23}$ & 52.0 \\
C$_2$H & $N=8-7, \; J=17/2-15/2$ & 698.545 & 429.17 & $1.68 \times 10^{-21}$ & $4.33 \times 10^{-23}$ & 38.9 \\
C$_2$H & $N=9-8, \; J=19/2-17/2$ & 785.802 & 381.51 & $1.88 \times 10^{-23}$ & $1.55 \times 10^{-24}$ & 12.2 \\
CN & $N=1-0, \; J=3/2-1/2$ & 113.495 & 2641.46 & $2.55 \times 10^{-21}$ & $1.38 \times 10^{-22}$ & 18.5 \\
CN & $N=2-1, \; J=5/2-3/2$ & 226.876 & 1321.39 & $3.66 \times 10^{-20}$ & $2.21 \times 10^{-21}$ & 16.5 \\
CN & $N=3-2, \; J=7/2-5/2$ & 340.249 & 881.10 & $1.22 \times 10^{-19}$ & $8.55 \times 10^{-21}$ & 14.2 \\
CN & $N=4-3, \; J=9/2-7/2$ & 453.607 & 660.91 & $2.18 \times 10^{-19}$ & $1.81 \times 10^{-20}$ & 12.1 \\
CN & $N=5-4, \; J=11/2-9/2$ & 566.947 & 528.78 & $2.79 \times 10^{-19}$ & $2.59 \times 10^{-20}$ & 10.8 \\
CN & $N=6-5, \; J=13/2-11/2$ & 680.264 & 440.70 & $2.93 \times 10^{-19}$ & $2.65 \times 10^{-20}$ & 11.1 \\
CN & $N=7-6, \; J=15/2-13/2$ & 793.554 & 377.78 & $2.60 \times 10^{-19}$ & $1.98 \times 10^{-20}$ & 13.2 \\
CN & $N=8-7, \; J=17/2-15/2$ & 906.811 & 330.60 & $1.78 \times 10^{-19}$ & $1.13 \times 10^{-20}$ & 15.8 \\
CO & $J=1-0$ & 115.271 & 2600.76 & $6.53 \times 10^{-21}$ & $1.02 \times 10^{-21}$ & 6.4 \\
CO & $J=2-1$ & 230.538 & 1300.40 & $8.73 \times 10^{-20}$ & $7.88 \times 10^{-21}$ & 11.1 \\

CO & $J=3-2$ & 345.796 & 866.96 & $4.63 \times 10^{-19}$ & $2.47 \times 10^{-20}$ & 18.8 \\
CO & $J=4-3$ & 461.041 & 650.25 & $1.55 \times 10^{-18}$ & $5.31 \times 10^{-20}$ & 29.2 \\
CO & $J=5-4$ & 576.268 & 520.23 & $3.89 \times 10^{-18}$ & $9.29 \times 10^{-20}$ & 41.8 \\
CO & $J=6-5$ & 691.473 & 433.56 & $8.02 \times 10^{-18}$ & $1.43 \times 10^{-19}$ & 55.9 \\
CO & $J=7-6$ & 806.651 & 371.65 & $1.44 \times 10^{-17}$ & $2.05 \times 10^{-19}$ & 70.1 \\
CO & $J=8-7$ & 921.780 & 325.23 & $2.32 \times 10^{-17}$ & $2.78 \times 10^{-19}$ & 83.4 \\
CO & $J=9-8$ & 1036.912 & 289.12 & $3.43 \times 10^{-17}$ & $3.41 \times 10^{-19}$ & 100.5 \\
CO & $J=10-9$ & 1151.985 & 260.24 & $4.74 \times 10^{-17}$ & $3.22 \times 10^{-19}$ & 147.2 \\
CO & $J=11-10$ & 1267.014 & 236.61 & $6.25 \times 10^{-17}$ & $2.09 \times 10^{-19}$ & 298.7 \\
CO & $J=12-11$ & 1381.995 & 216.93 & $7.81 \times 10^{-17}$ & $1.16 \times 10^{-19}$ & 671.8 \\
CO & $J=13-12$ & 1496.923 & 200.27 & $9.15 \times 10^{-17}$ & $6.86 \times 10^{-20}$ & 1334.7 \\
CO & $J=14-13$ & 1611.793 & 186.00 & $1.00 \times 10^{-16}$ & $4.60 \times 10^{-20}$ & 2182.3 \\
CO & $J=15-14$ & 1726.602 & 173.63 & $1.04 \times 10^{-16}$ & $3.51 \times 10^{-20}$ & 2967.9 \\
CO & $J=16-15$ & 1841.346 & 162.81 & $1.02 \times 10^{-16}$ & $2.90 \times 10^{-20}$ & 3519.6 \\
CO & $J=17-16$ & 1956.018 & 153.27 & $9.47 \times 10^{-17}$ & $2.46 \times 10^{-20}$ & 3856.1 \\
CO & $J=18-17$ & 2070.616 & 144.78 & $8.23 \times 10^{-17}$ & $1.94 \times 10^{-20}$ & 4242.4 \\
CO & $J=19-18$ & 2185.135 & 137.20 & $6.73 \times 10^{-17}$ & $1.34 \times 10^{-20}$ & 5034.3 \\
CO & $J=20-19$ & 2299.570 & 130.37 & $5.18 \times 10^{-17}$ & $8.11 \times 10^{-21}$ & 6385.6 \\
CO & $J=21-20$ & 2413.917 & 124.19 & $3.77 \times 10^{-17}$ & $4.48 \times 10^{-21}$ & 8415.6 \\
CO & $J=22-21$ & 2528.172 & 118.58 & $2.63 \times 10^{-17}$ & $2.32 \times 10^{-21}$ & $1.134 \times 10^{4}$ \\
CO & $J=23-22$ & 2642.330 & 113.46 & $1.76 \times 10^{-17}$ & $1.14 \times 10^{-21}$ & $1.552 \times 10^{4}$ \\
CO & $J=24-23$ & 2756.388 & 108.76 & $1.15 \times 10^{-17}$ & $5.34 \times 10^{-22}$ & $2.150 \times 10^{4}$ \\
CO & $J=25-24$ & 2870.339 & 104.44 & $7.28 \times 10^{-18}$ & $2.42 \times 10^{-22}$ & $3.006 \times 10^{4}$ \\
CO & $J=26-25$ & 2984.181 & 100.46 & $4.50 \times 10^{-18}$ & $1.06 \times 10^{-22}$ & $4.236 \times 10^{4}$ \\
CO & $J=27-26$ & 3097.909 & 96.77 & $2.73 \times 10^{-18}$ & $4.53 \times 10^{-23}$ & $6.031 \times 10^{4}$ \\
CO & $J=28-27$ & 3211.519 & 93.35 & $1.63 \times 10^{-18}$ & $1.88 \times 10^{-23}$ & $8.683 \times 10^{4}$ \\
CO & $J=29-28$ & 3325.005 & 90.16 & $9.58 \times 10^{-19}$ & $7.55 \times 10^{-24}$ & $1.269 \times 10^{5}$ \\
CO & $J=30-29$ & 3438.365 & 87.19 & $5.55 \times 10^{-19}$ & $2.96 \times 10^{-24}$ & $1.878 \times 10^{5}$ \\
CO & $J=31-30$ & 3551.592 & 84.41 & $3.16 \times 10^{-19}$ & $1.12 \times 10^{-24}$ & $2.813 \times 10^{5}$ \\
o-H$_2$CO & $J_{K_a, K_c} = 6_{1,5} - 6_{1,6}$ & 101.333 & 2958.49 & $7.91 \times 10^{-26}$ & $5.11 \times 10^{-25}$ & 0.2 \\
o-H$_2$CO & $J_{K_a, K_c} = 2_{1,1} - 1_{1,0}$ & 150.498 & 1992.00 & $2.13 \times 10^{-23}$ & $7.33 \times 10^{-23}$ & 0.3 \\
o-H$_2$CO & $J_{K_a, K_c} = 8_{1,7} - 8_{1,8}$ & 173.462 & 1728.29 & $7.35 \times 10^{-26}$ & $7.26 \times 10^{-25}$ & 0.1 \\
o-H$_2$CO & $J_{K_a, K_c} = 3_{1,2} - 2_{1,1}$ & 225.698 & 1328.29 & $1.31 \times 10^{-22}$ & $4.60 \times 10^{-22}$ & 0.3 \\
o-H$_2$CO & $J_{K_a, K_c} = 5_{3,3} - 4_{3,2}$ & 364.275 & 822.98 & $3.62 \times 10^{-23}$ & $1.21 \times 10^{-22}$ & 0.3 \\
o-H$_2$CO & $J_{K_a, K_c} = 7_{1,7} - 6_{1,6}$ & 491.968 & 609.37 & $5.15 \times 10^{-22}$ & $3.23 \times 10^{-21}$ & 0.2 \\
o-H$_2$CO & $J_{K_a, K_c} = 10_{1,10} - 9_{1,9}$ & 701.370 & 427.44 & $1.33 \times 10^{-22}$ & $1.52 \times 10^{-21}$ & 0.1 \\
o-H$_2$CO & $J_{K_a, K_c} = 12_{1,11} - 11_{1,10}$ & 896.805 & 334.29 & $1.42 \times 10^{-23}$ & $3.29 \times 10^{-22}$ & 0.0 \\
p-H$_2$CO & $J_{K_a, K_c} = 2_{0.2} - 1_{0,1}$ & 145.603 & 2058.97 & $1.53 \times 10^{-23}$ & $5.28 \times 10^{-23}$ & 0.3 \\
p-H$_2$CO & $J_{K_a, K_c} = 3_{0,3} - 2_{0,2}$ & 218.222 & 1373.79 & $8.15 \times 10^{-23}$ & $2.84 \times 10^{-22}$ & 0.3 \\
p-H$_2$CO & $J_{K_a, K_c} = 5_{2,3} - 4_{2,2}$ & 365.363 & 820.53 & $7.32 \times 10^{-23}$ & $3.36 \times 10^{-22}$ & 0.2 \\
p-H$_2$CO & $J_{K_a, K_c} = 6_{0,6} - 5_{0,5}$ & 434.493 & 689.98 & $3.48 \times 10^{-22}$ & $1.91 \times 10^{-21}$ & 0.2 \\
p-H$_2$CO & $J_{K_a, K_c} = 10_{0,10} - 9_{0,9}$ & 716.939 & 418.16 & $6.97 \times 10^{-23}$ & $8.49 \times 10^{-22}$ & 0.1 \\
p-H$_2$CO & $J_{K_a, K_c} = 12_{0,12} - 11_{0,11}$ & 855.151 & 350.57 & $1.20 \times 10^{-23}$ & $2.21 \times 10^{-22}$ & 0.1 \\
o-H$_2$O & $J_{K_a, K_c} = 2_{2,1} - 2_{1,2}$ & 1661.008 & 180.49 & $1.63 \times 10^{-19}$ & $3.73 \times 10^{-19}$ & 0.4 \\
o-H$_2$O & $J_{K_a, K_c} = 2_{1,2} - 1_{0,1}$ & 1669.905 & 179.53 & $1.37 \times 10^{-18}$ & $7.76 \times 10^{-19}$ & 1.8 \\
o-H$_2$O & $J_{K_a, K_c} = 3_{0,3} - 2_{1,2}$ & 1716.770 & 174.63 & $5.76 \times 10^{-19}$ & $5.57 \times 10^{-19}$ & 1.0 \\
o-H$_2$O & $J_{K_a, K_c} = 4_{1,4} - 3_{0,3}$ & 2640.474 & 113.54 & $3.71 \times 10^{-19}$ & $2.03 \times 10^{-19}$ & 1.8 \\
o-H$_2$O & $J_{K_a, K_c} = 2_{2,1} - 1_{1,0}$ & 2773.977 & 108.07 & $5.63 \times 10^{-19}$ & $8.00 \times 10^{-19}$ & 0.7 \\
o-H$_2$O & $J_{K_a, K_c} = 6_{1,6} - 5_{0,5}$ & 3654.604 & 82.03 & $7.22 \times 10^{-20}$ & $4.52 \times 10^{-20}$ & 1.6 \\
o-H$_2$O & $J_{K_a, K_c} = 4_{2,3} - 3_{1,2}$ & 3807.259 & 78.74 & $1.68 \times 10^{-19}$ & $1.01 \times 10^{-19}$ & 1.7 \\
o-H$_2$O & $J_{K_a, K_c} = 7_{10,7} - 6_{1,6}$ & 4166.852 & 71.95 & $3.15 \times 10^{-20}$ & $2.02 \times 10^{-20}$ & 1.6 \\
o-H$_2$O & $J_{K_a, K_c} = 8_{1,8} - 7_{0,7}$ & 4734.296 & 63.32 & $1.10 \times 10^{-20}$ & $4.17 \times 10^{-21}$ & 2.6 \\
p-H$_2$O & $J_{K_a, K_c} = 3_{1,3} - 2_{0,2}$ & 2164.132 & 138.53 & $3.32 \times 10^{-19}$ & $5.11 \times 10^{-19}$ & 0.6 \\
p-H$_2$O & $J_{K_a, K_c} = 4_{0,4} - 3_{1,3}$ & 2391.573 & 125.35 & $9.59 \times 10^{-20}$ & $4.97 \times 10^{-20}$ & 1.9 \\
HCN & $J=1-0$ & 88.632 & 3382.46 & $5.20 \times 10^{-22}$ & $6.46 \times 10^{-23}$ & 8.0 \\
HCN & $J=2-1$ & 177.261 & 1691.25 & $8.78 \times 10^{-21}$ & $1.13 \times 10^{-21}$ & 7.8 \\
HCN & $J=3-2$ & 265.886 & 1127.52 & $2.79 \times 10^{-20}$ & $4.62 \times 10^{-21}$ & 6.0 \\
HCN & $J=4-3$ & 354.505 & 845.66 & $4.57 \times 10^{-20}$ & $1.07 \times 10^{-20}$ & 4.3 \\
HCN & $J=5-4$ & 443.116 & 676.56 & $5.56 \times 10^{-20}$ & $1.76 \times 10^{-20}$ & 3.2 \\
HCN & $J=6-5$ & 531.716 & 563.82 & $5.55 \times 10^{-20}$ & $2.19 \times 10^{-20}$ & 2.5 \\
HCN & $J=7-6$ & 620.304 & 483.30 & $4.90 \times 10^{-20}$ & $2.06 \times 10^{-20}$ & 2.4 \\
HCN & $J=8-7$ & 708.877 & 422.91 & $4.04 \times 10^{-20}$ & $1.48 \times 10^{-20}$ & 2.7 \\
HCN & $J=9-8$ & 797.433 & 375.95 & $3.33 \times 10^{-20}$ & $8.98 \times 10^{-21}$ & 3.7 \\
HCN & $J=10-9$ & 885.970 & 338.38 & $2.77 \times 10^{-20}$ & $5.60 \times 10^{-21}$ & 5.0 \\
HCO$^+$ & $J=1-0$ & 89.189 & 3361.33 & $4.69 \times 10^{-21}$ & $5.52 \times 10^{-22}$ & 8.5 \\
HCO$^+$ & $J=2-1$ & 178.375 & 1680.68 & $5.69 \times 10^{-20}$ & $4.34 \times 10^{-21}$ & 13.1 \\
HCO$^+$ & $J=3-2$ & 267.558 & 1120.48 & $2.17 \times 10^{-19}$ & $1.37 \times 10^{-20}$ & 15.9 \\
HCO$^+$ & $J=4-3$ & 356.734 & 840.38 & $5.17 \times 10^{-19}$ & $2.99 \times 10^{-20}$ & 17.3 \\
HCO$^+$ & $J=5-4$ & 445.903 & 672.33 & $9.96 \times 10^{-19}$ & $5.28 \times 10^{-20}$ & 18.9 \\
HCO$^+$ & $J=6-5$ & 535.062 & 560.30 & $1.68 \times 10^{-18}$ & $8.13 \times 10^{-20}$ & 20.6 \\
HCO$^+$ & $J=7-6$ & 624.209 & 480.28 & $2.52 \times 10^{-18}$ & $1.14 \times 10^{-19}$ & 22.1 \\
HCO$^+$ & $J=8-7$ & 713.342 & 420.26 & $3.45 \times 10^{-18}$ & $1.54 \times 10^{-19}$ & 22.5 \\
HCO$^+$ & $J=9-8$ & 802.458 & 373.59 & $4.33 \times 10^{-18}$ & $1.95 \times 10^{-19}$ & 22.2 \\
HCO$^+$ & $J=10-9$ & 891.558 & 336.26 & $4.98 \times 10^{-18}$ & $2.16 \times 10^{-19}$ & 23.0 \\
$[$OI$]$ & $^3P_1-^3P_2$ & 2061.069 & 145.45 & $4.06 \times 10^{-16}$ & $4.21 \times 10^{-19}$ & 964.4 \\
$[$OI$]$ & $^3P_0-^3P_1$ & 4747.777 & 63.14 & $8.05 \times 10^{-15}$ & $6.49 \times 10^{-18}$ & 1239.9 \\
OH & $F=3/2^+ - 1/2^-$ & 1834.760 & 163.40 & $3.21 \times 10^{-18}$ & $5.58 \times 10^{-20}$ & 57.5 \\
OH & $F=3/2^+ -1/2^-$ & 1839.047 & 162.02 & $4.22 \times 10^{-18}$ & $4.93 \times 10^{-20}$ & 85.6 \\
OH & $F=5/2^+ -3/2^-$ & 2509.952 & 119.44 & $2.34 \times 10^{-17}$ & $1.20 \times 10^{-18}$ & 19.6 \\
OH & $F=5/2^- -3/2^+$ & 2514.312 & 119.23 & $1.56 \times 10^{-17}$ & $1.07 \times 10^{-18}$ & 14.6 \\
OH & $F=7/2^- -5/2^+$ & 3543.787 & 84.60 & $1.34 \times 10^{-17}$ & $3.03 \times 10^{-19}$ & 44.1 \\
OH & $F=7/2^+ -5/2^-$ & 3551.192 & 84.42 & $4.64 \times 10^{-18}$ & $2.56 \times 10^{-19}$ & 18.1 \\
OH & $F=1/2^- -3/2^+$ & 3786.259 & 79.18 & $2.05 \times 10^{-17}$ & $3.60 \times 10^{-19}$ & 57.0 \\
OH & $F=9/2^+ -7/2^-$ & 4592.491 & 65.28 & $3.29 \times 10^{-18}$ & $7.51 \times 10^{-20}$ & 43.8 \\
OH & $F=9/2^- -7/2^+$ & 4602.864 & 65.13 & $6.79 \times 10^{-19}$ & $7.05 \times 10^{-20}$ & 9.6 \\
\end{longtable}
\end{small}

\twocolumn 

\section{Line flux contribution functions}
\begin{figure*}
\centering
\includegraphics[clip=,width=1.0\linewidth]{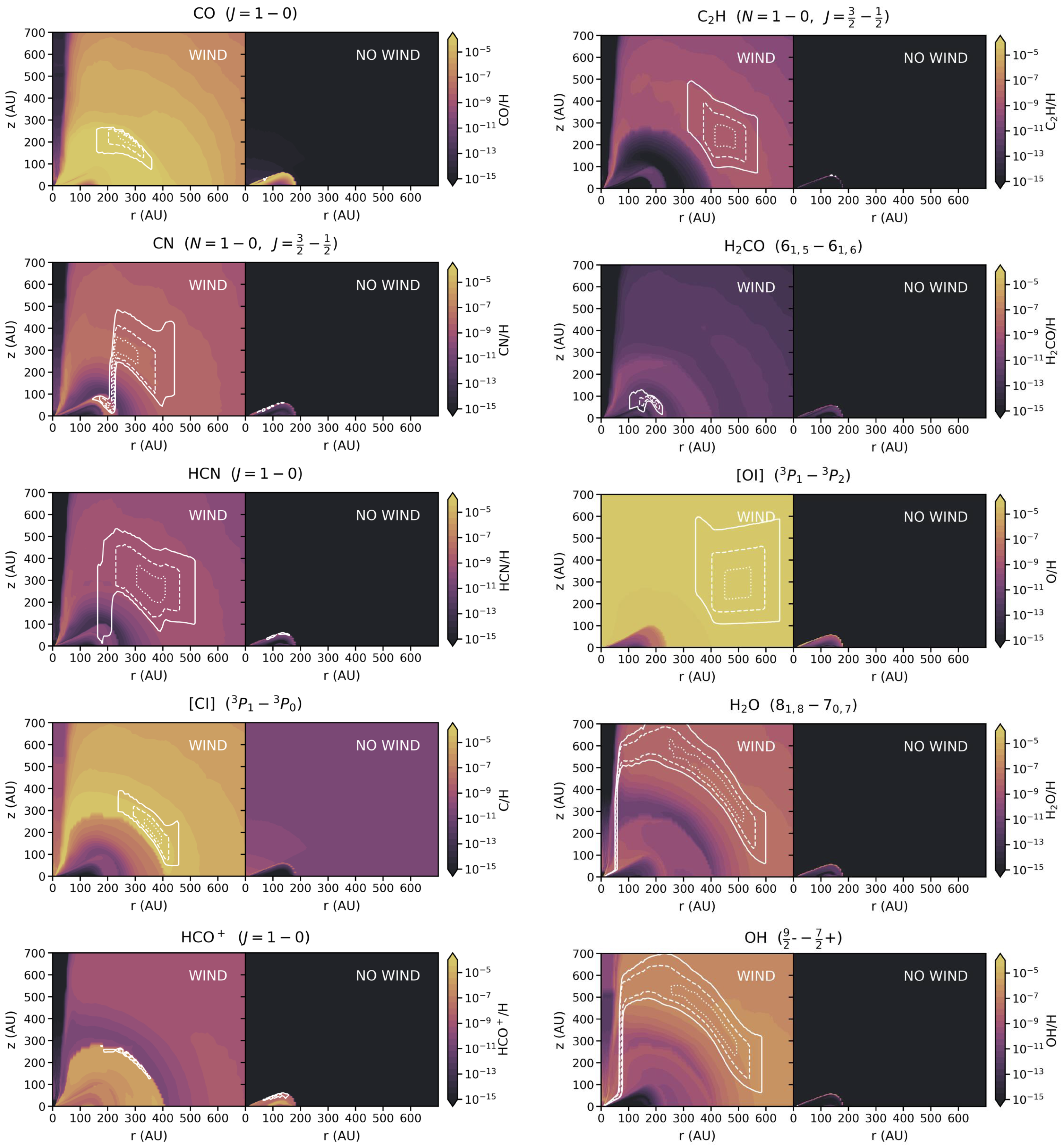}
\caption{Contribution functions for 10 representative transitions of 10 different species, comparing full-wind and no-wind models. The colormaps show the abundance of a given species and the white lines denote the 40\%, 60\%, and 80\% emission contours of a given transition. In the full-wind model, the disk-integrated line fluxes are dominated by emission from the wind itself rather than from the disk. Note that in no-wind cases where the contribution function is not visible, emission arises predominantly from a small region in the inner disk that is not captured at these scales.}
\label{fig_cbfs}
\end{figure*}

\section{Model parameter exploration}
\label{apx:model_param_exploration}
\begin{figure*}
\centering
\includegraphics[clip=,width=1.0\linewidth]{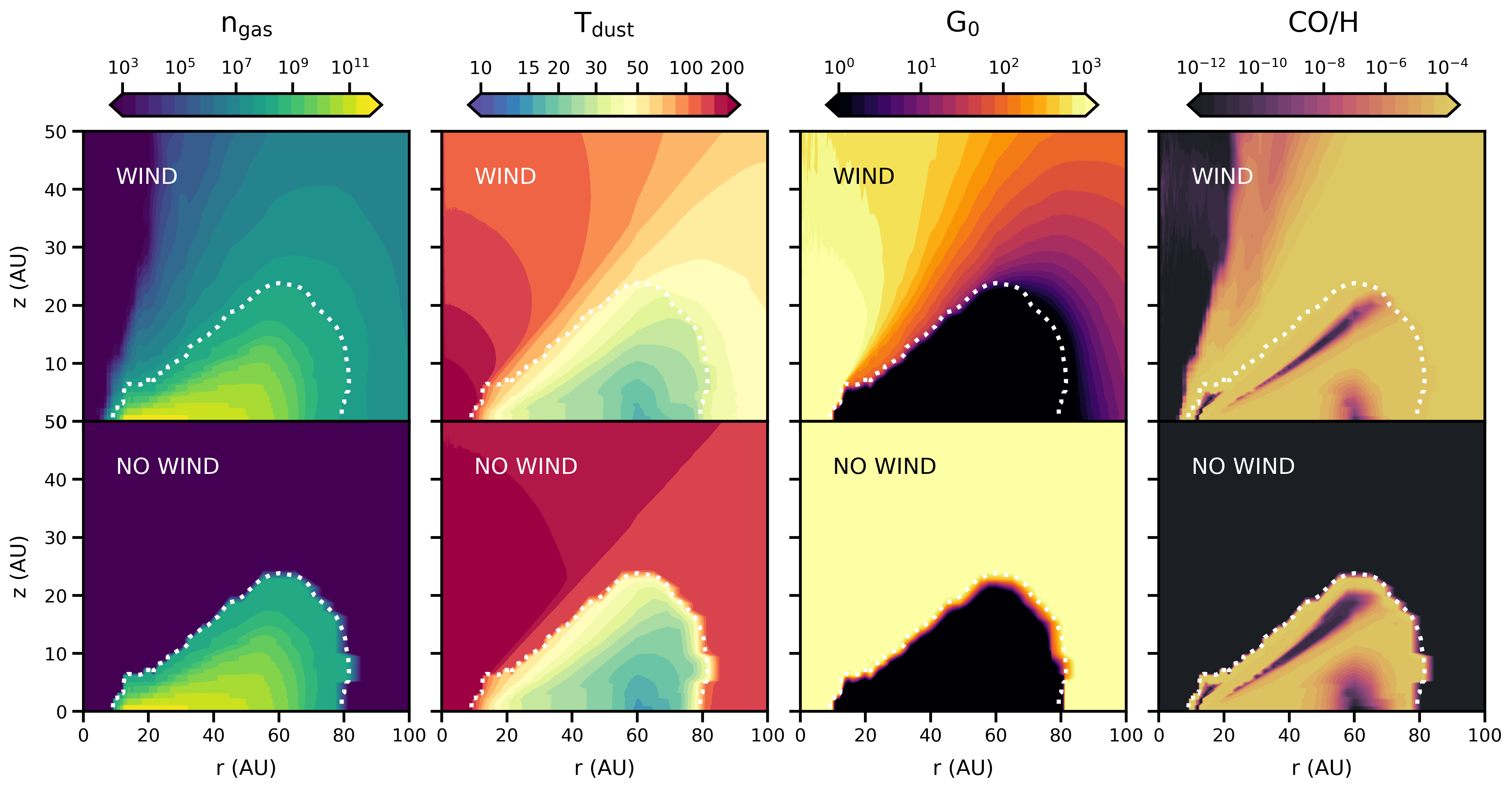}
\caption{Properties of the model where the disk size is initialised at $r_\text{d} = 50$ au. Models including the wind are shown in the top row, and models in which the wind has been artificially removed are shown in the bottom row. \emph{First column: }Gas number density. \emph{Second column: }Dust temperature. \emph{Third column: }External integrated UV radiation field strength (in Habing units G$_0$). \emph{Fourth column: }Gas-phase CO abundance.}
\label{fig_50au_disk_model}
\end{figure*}
\begin{figure}
\centering
\includegraphics[clip=,width=1.0\linewidth]{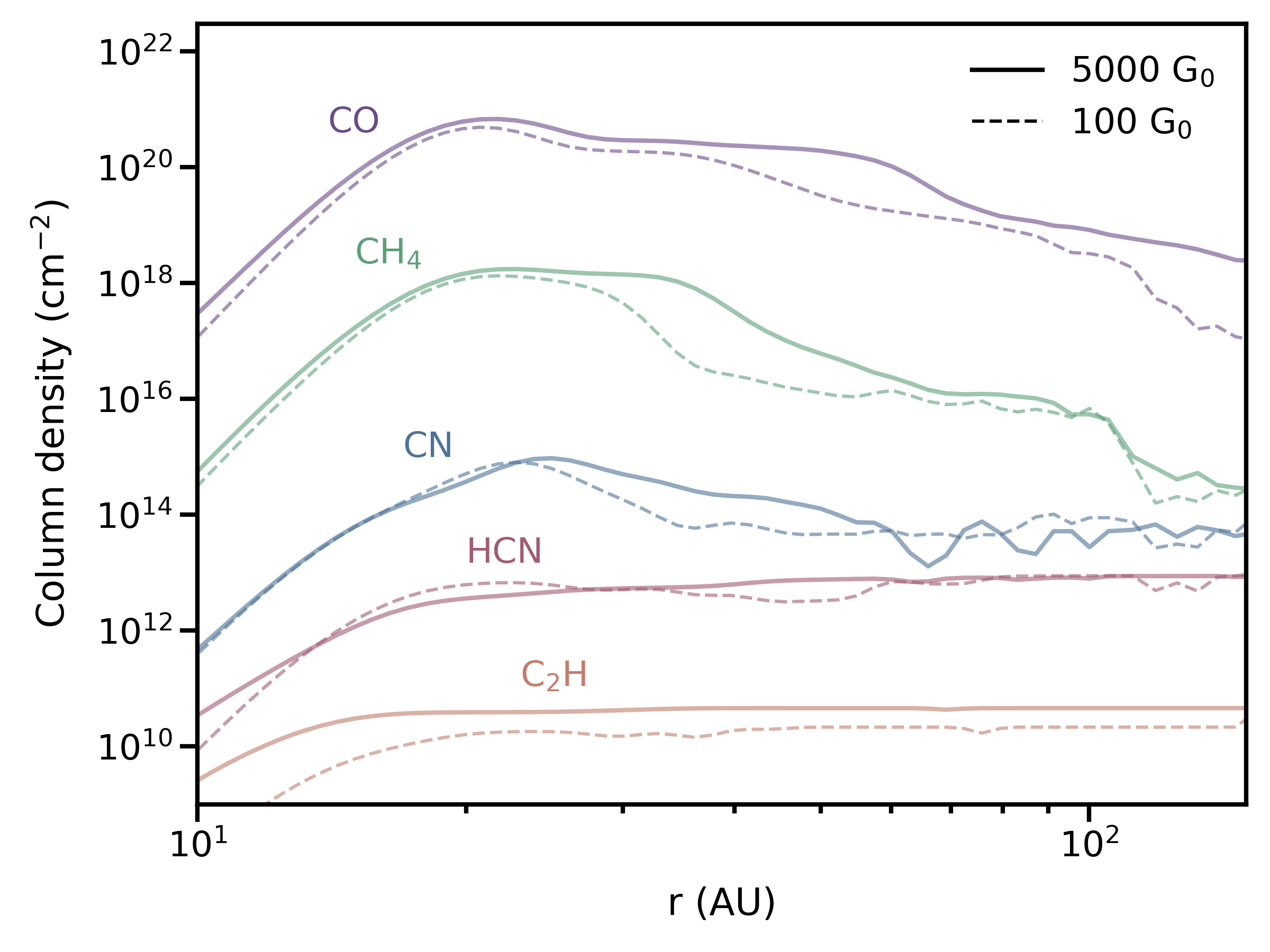}
\caption{Column densities for several key species, extracted from windless models that have external UV fields of 100 G$_0$ (dashed lines) and 5000 G$_0$ (solid lines). The column densities are typical enhanced by a factor of a few for the higher strength external UV field, in agreement with \citet{walsh_2013}.}
\label{fig_column_densities}
\end{figure}

\section{Reaction rates analysis}
\label{apx:reaction_rates_analysis}

In Section \ref{subsec:results_composition}, we showed that differences in the abundances of key volatiles (CO, H$_2$O, and CH$_4$) when comparing the `wind' and `no wind' models are primarily driven by the balance between desorption and freeze-out. To demonstrate this quantitatively, we present here a detailed analysis of the reactions for CH$_4$, chosen as a representative example to illustrate the relative importance of various formation and destruction pathways.

\begin{figure*}
\centering
\includegraphics[clip=,width=1\linewidth]{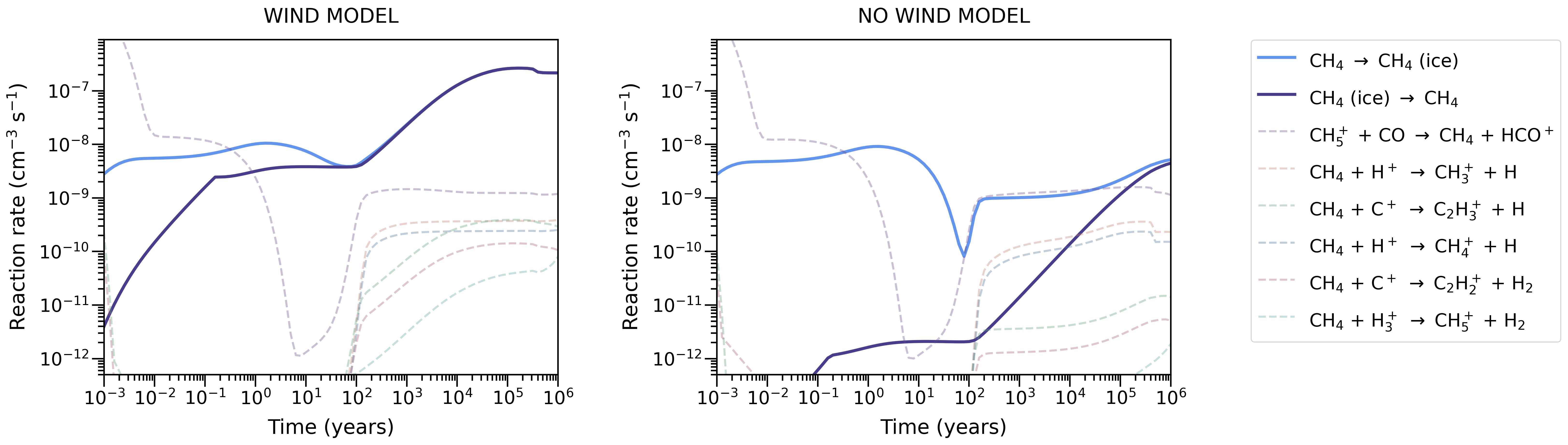}
\caption{CH$_4$ formation and destruction rates extracted from our `wind' (left) and `no wind' (right) models. In both cases, the eight reactions with the highest rates are shown, as measured at the end of the model run ($t=1$ Myr). The key processes that regulate the gas-phase CH$_4$ abundance in both cases are thermal desorption (solid dark blue line) and freezeout (solid light blue line). Gas-phase formation and destruction pathways (dotted lines) are comparatively less important.}
\label{fig_ch4_reactions}
\end{figure*}

Figure \ref{fig_ch4_reactions} shows the temporal evolution of reaction rates for the eight most significant CH$_4$-related processes (as measured at the end of the model run, $t=1$ Myr). These rates are extracted from a disk location of $r=120$ au and $z=50$ au, where abundance variations are highly pronounced (matching our analysis in Section \ref{subsec:results_composition}). The two models display distinct variations over the 1 Myr timescale.

In the `wind' model, the CH$_4$ thermal desorption and freeze-out rates (solid dark and light blue lines respectively) become balanced relatively quickly (reaching the same value after $\sim 10^2$ years, then continuing to rise in tandem until plateauing at a value of $\sim 3 \times 10^{-7}$ cm$^{-3}$ s$^{-1}$). In contrast, the `no wind' model shows a marked imbalance; the thermal desorption rate remains orders of magnitude below the freezeout rate for most of the simulation, achieving equilibrium only at $t \sim 1$ Myr with a significantly lower rate of $\sim 4 \times 10^{-9}$ cm$^{-3}$ s$^{-1}$. This disparity stems from the temperature dependence of the thermal desorption rate (implemented following \citealt{visser_2011}):

\begin{gather}
R_{\text{thdes}}(X) = 4\pi a_\text{gr}^2 n_\text{gr} f(X) \nu(X) N_\text{ss} \exp{\bigg[-\frac{E_\text{b}(X)}{kT_\text{dust}}\bigg]}
\label{eq:thermal_desorption}
\end{gather}
where $a_\text{gr} = 0.1 \mu$m is the typical grain radius, $n_\text{gr} = 10^{-12} n_\text{gas}$ is the grain number density, $f$ is a dimensionless factor that each species desorbs according to its solid-phase abundance, $\nu=2\times 10^{12}$ s$^{-1}$ is the canonical pre-exponential factor for first-order desorption \citep{sandford_allamandola_1993}, $N_\text{ss} = 8 \times 10^{14}$ cm$^{-2}$ is the number of binding sites per unit grain surface, $E_\text{b}$ is the binding energy, and $k$ is the Boltzmann constant. At our analysed location, the wind model maintains a dust temperature approximately 4 K higher than the no-wind model.

The gas-phase chemistry, while exhibiting complex temporal behaviour, plays a secondary role in determining CH$_4$ abundance. The primary formation pathway in both models:
\begin{gather}
\text{CH$_5^+$ + CO} \rightarrow \text{CH$_4$ + HCO$^+$}
\label{eq:ch4_reaction_1}
\end{gather}
maintains a relatively stable rate of $\sim10^{-9}$ cm$^{-3}$ s$^{-1}$ after $\sim 10^2$ years in both models. The remaining gas-phase processes consist of ion-molecule destruction reactions:
\begin{gather}
\text{CH$_4$ + H$^+$} \rightarrow \text{CH$_3^+$ + H$_2$} \\
\text{CH$_4$ + C$^+$} \rightarrow \text{C$_2$H$_3^+$ + H} \\
\text{CH$_4$ + H$^+$} \rightarrow \text{CH$_4^+$ + H} \\
\text{CH$_4$ + C$^+$} \rightarrow \text{C$_2$H$_2^+$ + H$_2$} \\
\text{CH$_4$ + H$_3^+$} \rightarrow \text{CH$_5^+$ + H$_2$}
\label{eq:ch4_reaction_2}
\end{gather}
These destruction pathways become active after $\sim 10^2$ years and eventually stabilize at rates between $\sim 10^{-12}$ and $10^{-10}$ cm$^{-3}$ s$^{-1}$. Importantly, their maximum rates remain 1-3 orders of magnitude below the freeze-out/desorption rates throughout the simulation in both models. Therefore, while the wind affects multiple aspects of CH$_4$ chemistry, the abundance patterns are predominantly determined by the balance between freeze-out and desorption rather than gas-phase reactions.

It is important to note that this detailed reaction analysis applies specifically to CH$_4$ our chosen disk location ($r=120$ au, $z=50$ au). While our broader findings about the dominance of desorption/freeze-out processes over gas-phase chemistry hold true across much of the disk, the specific gas-phase reactions described here represents just one regional example. Different disk locations, characterized by varying physical conditions (density, temperature, radiation field), will feature distinct sets of gas-phase reactions. The ion-molecule pathways identified in our analysis, though secondary to the desorption/freezeout processes in determining CH$_4$ abundance, are particular to the physical conditions at this location. A complete understanding of CH$_4$ chemistry throughout the disk would require analysing how these reaction networks vary across different disk regions, and will be addressed in future studies.

\bsp	
\label{lastpage}
\end{document}